\documentclass[fleqn,usenatbib]{mnras}

\usepackage[T1]{fontenc}

\DeclareRobustCommand{\VAN}[3]{#2}
\let\VANthebibliography\thebibliography
\def\thebibliography{\DeclareRobustCommand{\VAN}[3]{##3}\VANthebibliography}

\usepackage{graphicx}	
\usepackage{amsmath}	
\usepackage{amssymb}	
\usepackage{caption}
\usepackage{subcaption}
\usepackage{soul}

\usepackage{multirow}
\usepackage{multicol}
\newcommand{\msol}{$M_{\odot}\,$}
\newcommand{\csec}{$\text{count}\,\text{s}^{-1}\,$}
\newcommand{\Mdot}{$\dot{M}\,$}
\newcommand{\chired}{$\chi_{\nu}^{2}\,$}
\newcommand{\chist}{$\chi^{2}\,$}

\newcommand{\unilum}{$\text{erg}/\text{s}\,$}

\title[LMC X-4 disk precession]{Disk precession to explain the super-orbital modulation of\\
LMC X-4: results from the \textit{Swift} monitoring campaign}
\author[Ambrosi E. et al.]{
E. Ambrosi$^{1}$\thanks{E-mail: elena.ambrosi@inaf.it},
A. D'Aì$^{1}$,
M. Del Santo$^{1}$,
A. Segreto$^{1}$, 
C. Ferrigno$^{2}$, 
R. Amato$^{3}$, 
\newauthor 
G. Cusumano$^{1}$\\
$^{1}$ INAF/IASF Palermo, via Ugo La Malfa 153, I-90146 - Palermo, Italy. \\
$^{2}$ISDC, Department of astronomy, University of Geneva, chemin d’Ecogia, 16; 1290 Versoix, Switzerland.\\
$^{3}$IRAP, Universit\'e de Toulouse, CNRS, UPS, CNES, Toulouse, France.\\
}

\date{Accepted XXX. Received YYY; in original form ZZZ}

\pubyear{}
\usepackage{newtxtext,newtxmath}
\begin{document}
\label{firstpage}
\pagerange{\pageref{firstpage}--\pageref{lastpage}}
\maketitle

\begin{abstract}
We studied the spectral changes of the high-mass X-ray binary system LMC X-4 to understand the origin and mechanisms beyond its super-orbital modulation (30.4 days). To this aim, we obtained a monitoring campaign with Swift/XRT (0.3-10 keV) and complemented these data with the years-long Swift/BAT survey data (15-60 keV). We found a self-consistent, physically motivated, description of the broadband X-ray spectrum using a Swift/XRT and a NuSTAR observation at the epoch of maximum flux. We decomposed the spectrum into the sum of a bulk+thermal Comptonization, a disk-reflection component and a soft contribution from a standard Shakura-Sunyaev accretion disk. 
We applied this model to 20 phase-selected \textit{Swift} spectra along the super-orbital period. We  found a phase-dependent flux ratio of the different components, whereas the absorption column does not significantly vary. The disk emission is decoupled with respect to the hard flux. We interpret this as a geometrical effect in which the inner parts of the disk are tilted with respect to the \textit{obscuring} outer regions.
\end{abstract}




\section{Introduction}
High-Mass X-ray Binaries (HMXBs) are accreting binary systems where a compact object like a neutron star (NS) or a black hole (BH) accretes matter from a massive donor (i.e. O-B stars). The phenomenology of these type of binaries is driven by mass lost by the donor via stellar wind and/or Roche Lobe Overflow (RLO) which is captured by the gravitational field of the compact object. The majority of HMXBs show their flux modulated at their orbital period not only in the X-rays but also at other longer wavelengths \citep{2008MNRAS.387..439R,2009aaxo.conf...70Z,2006ApJS..163..372W}. Some HMXBs are characterized also by variations at longer time scales, the so-called super-orbital periods \citep{2012MNRAS.420.1575K}. Orbital periods ($P_{\text{orb}}$) are usually found in the range of some days up to dozens, while super-orbital modulations ($P_{\text{sup}}$) have time scales from tens to hundreds of days. 
The orbital modulations can  explained as due to occultation of the compact object by the massive companion. In other cases more complex processes intervene: the absorption and scattering of X-rays by the donor wind, or the deviation from the spherical shape of the Roche Lobe filling star, as in Cyg X-1 \citep[see][]{1999ApJ...525..968W,1999A&A...343..861B}, or 
the scattering in hot gas around a bulge at the disc edge
\citep[as in the case of 4U 1820--303,][]{zdziarski2007}. 
The origin of super-orbital modulation is instead explained invoking either radiation-induced disk warping \citep{2001MNRAS.320..485O}, or disk/jet precession \citep{1991MNRAS.249...25W}. 
Additional mechanisms that can lead to super-orbital modulations are: the magnetic warping \citep{2004ApJ...604..766P}, wind-driven tilting \citep{1994A&A...289..149S} and X-ray state changes due to mass transfer rate variations \citep{2006csxs.book..157M}. \citet{2012MNRAS.420.1575K} provides a comprehensive review of all these mechanisms.\\
In the present work we focus on the super-orbital variability of the HMXB LMC X-4.\\
\subsubsection*{The source LMC X-4}
LMC X-4 is an eclipsing HMXB located in the Large Magellanic Cloud (LMC) at a distance of $\sim$\,50 kpc \citep{1972ApJ...178..281G} and inclination $i \sim 59$\textdegree$\,$ \citep{2019PASJ...71...36I}. It is one of the most studied and well characterized binary systems, in which a 1.57 \msol NS, of $P_{s} = 13.5$ s \citep{1983ApJ...274..765K}, accretes persistently matter from its companion, a 18 \msol O7 III donor \citep{1978ApJ...225..548H}. The NS orbits the donor with an almost circular orbit (eccentricity $< 0.006$), with an orbital period of 1.4 d, which decays significantly: $\dot{P}/P = -1.00\times 10^{-6} \text{yr}^{-1} $ \citep{2015A&A...577A.130F}.
Long term X-ray monitoring revealed that its luminosity is persistently high ($L_{\mathrm{X}} \sim 2\times 10^{38}$ \unilum), with frequent flares reaching super-Eddington luminosity (a few $10^{39}$ \unilum). This HMXB has one of the most stable super-orbital modulation which does not show period shifts nor change in amplitude \citep{Molkov15}. Its super-orbital periodicity is about 30.4 d \citep{1981ApJ...246L..21L,1984A&A...140..251I,1989A&A...223..154H} and is observed in soft and hard X-rays, as well as in the UV and optical frequencies. The super-orbital X-ray flux variation is much larger than the orbital one: from the low state to the high state the intensity changes by a factor of $\sim 60$ \citep{2003A&A...401..265N}, while during the orbital eclipse the flux decreases by a factor of $\sim 2$ \citep{Molkov15}.\\
Numerous studies, both theoretical and observational, have been performed with the aim to understand the mechanisms behind the super-orbital variability of binaries. LMC X-4 is a benchmark for understanding the mechanisms at the origin of super-orbital flux modulation. Observational works analyzed and modelled the lightcurve modulation, the spectral characteristic as well as the spectroscopic signatures at different energies as function of super-orbital phase.
\citet{1989A&A...223..154H} reproduced the optical superorbital folded light curve with a geometric model which accounts for ellipsoidal variations, shielding of the emission of the donor due to orbital motion, under the assumption of a precessing accretion disk.
A further analysis with years-long X-ray coverage gave support to this interpretation. In fact later on, both \cite{2012MNRAS.420.1575K} and \cite{2019PASJ...71...36I} explained the modulation of the light curve in terms of disk precession, even if their two approaches were based on different assumptions, a point to which we will return in Sec. \ref{sec:discussion}.
Strong observational evidence has not yet been found for the origin of disk precession. \cite{2012MNRAS.420.1575K} argued in favour of a radiation-induced warping model \citep{2001MNRAS.320..485O} based on the stability of the light curve and the geometry of the binary. Further hints of disc precession have been found from pulse-phase spectroscopy. \citet{2004ApJ...600..351N} showed that the variation of the soft component over the pulsation period differs from that of the hard component.
This behaviour can be explained with the model developed by \cite{2005ApJ...633.1064H}, which computes the emission of the innermost regions of a precessing accretion disk irradiated by the NS beam. Using this model of disc precession, \citet{brumback2020} was able to explain the energy-resolved pulse profiles of this source.\\ Moreover, \cite{nielsen09_spec} investigated spectroscopic signatures of disk precession. Through a X-ray high-resolution analysis, they analysed a series of recombination X-ray emission lines at different super-orbital phases, located their origin and explained the relative variation in terms of precession of a warped accretion disk.  \\
In addition to the studies mentioned above, there are several investigations also on the spectral properties at different super-orbital stages. \cite{2003A&A...401..265N} analyzed low and high state spectra with RXTE/PCA observations, and found a variation on the photon index of the power-law component, with a flattening observed in the low state respect to the high state. Moreover, they found clear evidence of a positive correlation between the flux in the range 7-25 keV, and the flux of the iron emission line, whose equivalent width varies along the low state. This fact, they noted, interestingly may indicate the presence of a second region distant from the main X-ray source, where the iron line is produced. \cite{2010ApJ...720.1202H} analyzed the X-ray broad band (0.6-50 keV) behaviour of LMC X-4 in three different observations taken at specific flux levels that trace the long-term spectral trend of this source.
There, it has been found  that the SED can be well fitted with a multicomponent model, an absorbed soft black body with a high energy cutoff power-law, together with a number of emission lines, which do not seem to vary along the different phases. They also analyzed the pulse profile at low, medium and high energies of these three different super-orbital phases. The properties of the correlation function of the pulse profiles at different energies suggest that the cause of the soft spectrum can be the emission of reprocessed hard X-rays (produced in the innermost regions) from a precessing accretion disk.\\
All the above-mentioned studies have contributed to the understanding of super-orbital modulations on LMC X-4 and NS binaries in general, but they lacked a continuous coverage of broad-band X-ray data along the super-orbital period. In addition to this, phase-dependent studies have usually been performed using phenomenological models like the \textsc{npex} model \citep{Npex_mihara,brumback2020}, or a power-law \citep{nielsen09_spec}.\\
In this work we present an almost complete coverage of the $\sim30.4$ d super-orbital period both in the soft and hard X-ray bands. With the aim to understand how the observed luminosity and spectral shape vary on the super-orbital phase, we asked for a continuous monitoring campaign with the \textit{Neil Gehrels Swift Observatory} (\textit{Swift} hereafter, \citealt{2004ApJ...611.1005G}) along an entire cycle of the super-orbital period. We complemented the soft X-ray spectral coverage at higher energies with the super-orbital phase-resolved spectra obtained through $\approx 4.75$ years of the transient monitoring campaign of the BAT instrument aboard \textit{Swift}. We show the results of our spectral analysis using a self-consistent physical model which reproduces the spectrum from soft to hard X-rays. \\
\subsubsection*{Plan of the paper}
In Sec. 2 we introduce the data analyzed in this work and outline the reduction procedure. Then, in Sec. 3, we describe how the data are screened in order to obtained super-orbital phase-resolved spectra of both \textit{Swift} BAT and XRT instruments. Then in Sec. 4, we test simple models to the BAT and XRT spectra with the aim to describe the evolution of the spectral parameters along the super-orbital modulation. After this, we determine in Sec. 5 the best physical model that describes the broad-band data of the epoch of maximum flux, obtained with the Swift/XRT spectrum and a NuSTAR observation of high quality, and characterize the source properties. Finally, in Sec. 6 we apply the best physical model that describes the epoch of maximum flux to the super-orbital phase resolved XRT and BAT joint spectra. We discuss our results in Sec. \ref{sec:discussion}. 





\section{Observations and data reduction}
\label{sec:data}
%
\subsection{\textit{Swift}/XRT data}
\textit{Swift}/XRT pointed LMC X-4 37 times between 2019-05-24 and 2019-07-09 (ObsIDs: 00033538011-00033538048) covering about an entire super-orbital period, for a total exposure time of 101.8 ks. The data reduction has been performed with HEASoft (v. 6.27.2) and CALDB version 20200724. Clean and calibrated files were obtained using the task \textit{xrtpipeline}. We selected only grade 0 events for the following analysis.\\
%
Data taken with XRT in PC mode are affected by pile-up if the count rate from a source is $\gtrsim 0.5 $ \csec . We checked the presence of pile-up in each observation and corrected it accordingly to the standard pile-up mitigation procedure.\footnote{\url{https://www.swift.ac.uk/analysis/xrt/pileup.php}}
%
%
We then run the task \textit{xrtproducts} to extract the high level spectra and light curves. The background-subtracted light curves are generated with the \textsc{ftool} \textit{lcmath} adopting, for each IDs, the corresponding ratio of the source area to background area for the extraction regions. High level spectra are grouped with 20 counts per bin using the \textsc{grppha} task, in order to use the \chist statistic in the fitting procedure.
\begin{figure}
    \includegraphics[width=\columnwidth]{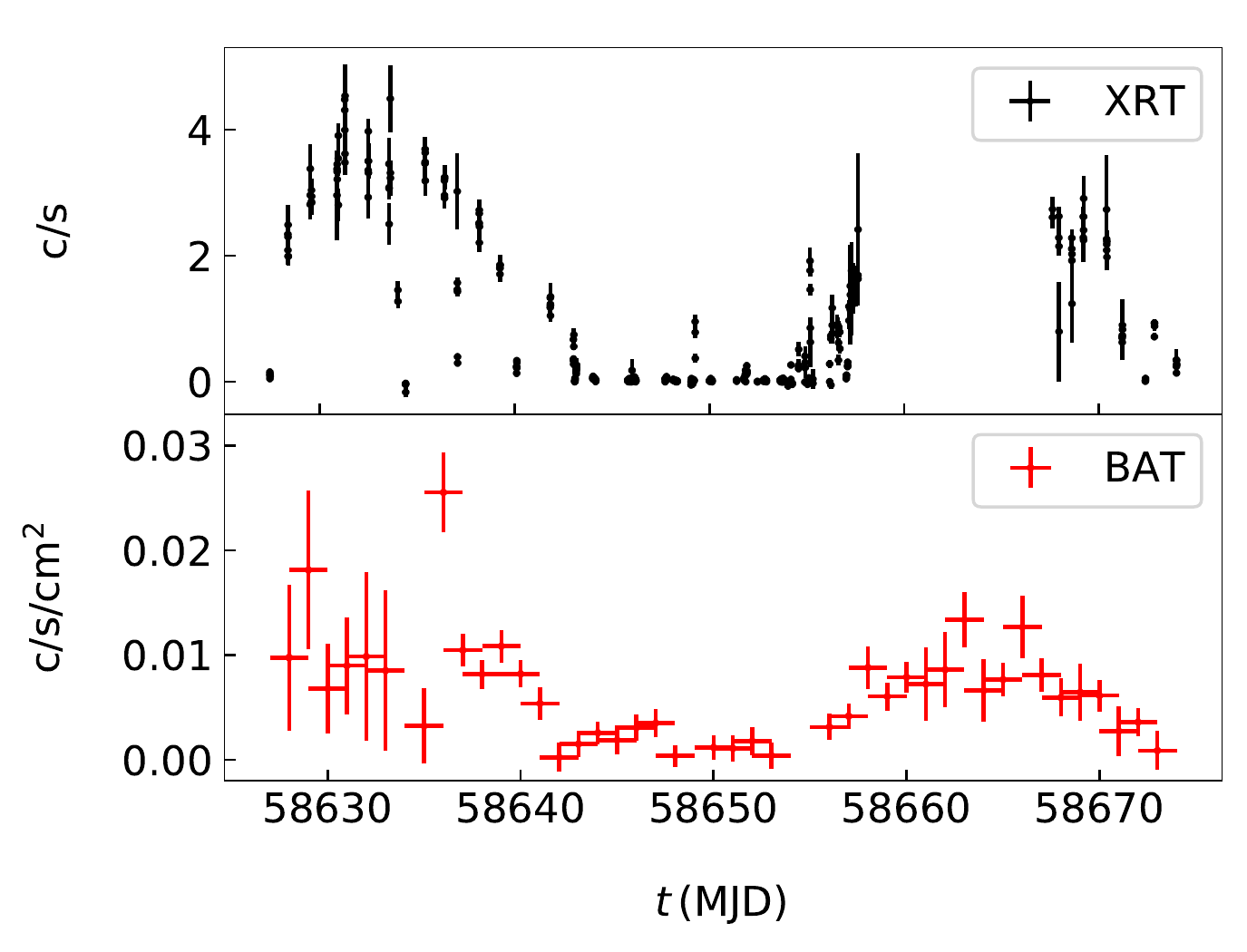}
    \caption{Upper panel: Swift/XRT light curve of the data collected between 2019-05-24 and 2019-07-09 with a time bin size of 300s. Bottom panel: Daily \textit{Swift}/BAT light curve in the 15-50 keV energy range collected over the same period of the XRT data. The two light curves clearly show the typical flux  modulation over the 30.4-day period.}
    \label{fig:xrt_total_lcurve}
\end{figure}
%
%
The upper panel of Fig.~\ref{fig:xrt_total_lcurve} shows the XRT light curve of our data-set with a bin-time of 300 s.
After visual inspection, we excluded from our analysis the ObsID 00033538041, because of the presence of a strong flare. 
\subsection{\textit{Swift}/BAT data}
\label{sub:bat_data}

\begin{figure}
    \centering
    \includegraphics[width=\columnwidth]{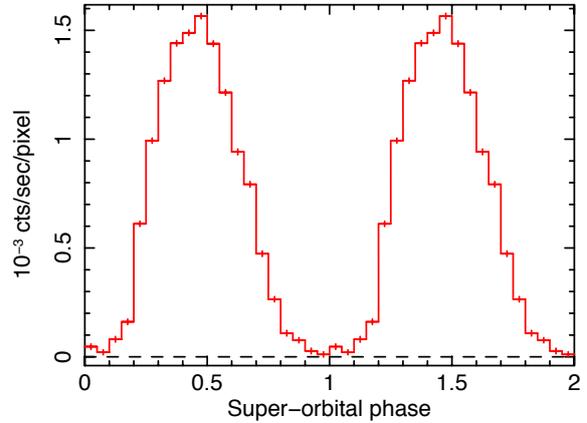}\caption{BAT folded light curve (with respect to P$_{\textrm {\rm sup}}$, divided into 20 bins. ($\psi_{\textrm {\rm sup}} =0.5$ corresponds to the period of maximum flux) in the energy range 15-50 keV.}
    \label{fig:lc_bat_folded}
\end{figure}

In the upper panel of Fig.~\ref{fig:xrt_total_lcurve} we show the Swift/BAT light curve (one-day averaged, 15-50 keV range) as obtained from the transient BAT monitor page\footnote{\url{https://swift.gsfc.nasa.gov/results/transients/LMCX-4/}} \citep{bat_monitor} during the super-orbital cycle examined in this work. The signal-to-noise ratio (SNR) for these BAT data is low, preventing simultaneous XRT-BAT spectral analysis.
To increase the SNR we stacked the high-energy BAT data from 2015 January 16 (MJD 57038) until 2019 October 30 (MJD 58776), and produced phase-resolved light curves and spectra. This choice is motivated by the observational evidence that the super-orbital period of LMC X-4 has been stable over more than twenty years \citep{Molkov15}.  Data were reprocessed using the \textsc{batimager} code  \citep{segreto_code}. \textsc{batimager}  performs image reconstruction via cross-correlation and generates light curves and spectra for each detected source. We extracted the super-orbital folded light curve and spectra, dividing the $P_{\mathrm{sup}}\sim30.4$ d in 20 equal phase-bins, where we set the epoch of maximum flux at $\psi_{\text{sup}}=0.5$. Each phase-bin lasts approximately one orbital period in order to smooth any orbital-dependent variations. The super-orbital phase-resolved lightcurve of the BAT data collected during $ \approx 4.75$ years of observations is shown in Fig. \ref{fig:lc_bat_folded}.
Spectra were extracted in the range 10-150 keV, with logarithmic binning for a total of 56 bins. However, the bins with energy higher than 60 keV were background dominated, so we analyzed the spectra in the range 15-60 keV. We used the official BAT spectral redistribution matrix.

\subsection{\textit{NuSTAR} data}
\begin{figure}
    \centering
    \includegraphics[width=\columnwidth]{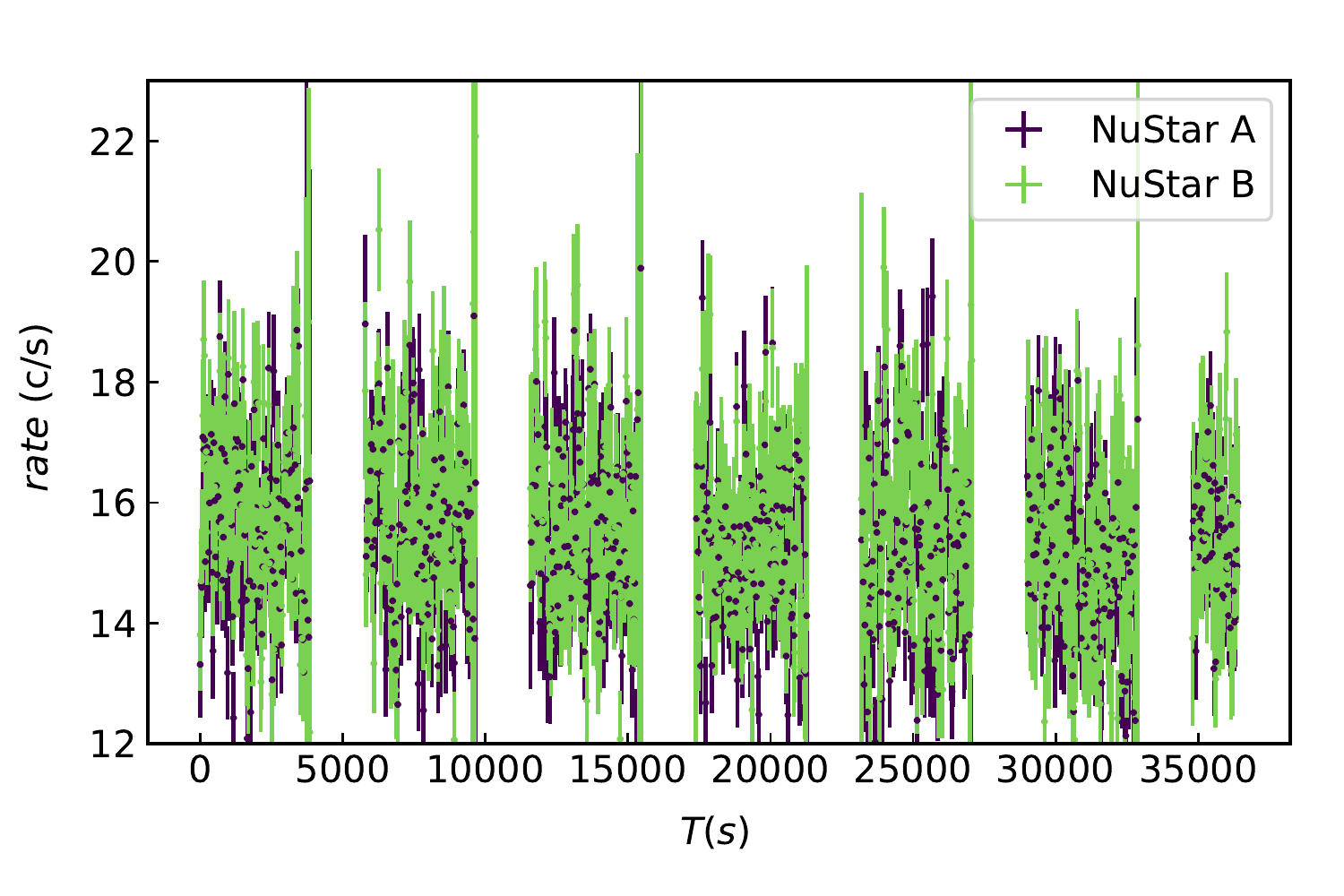}
    \caption{Light curve of the \textit{NuSTAR} data binned at 200 s. $T(s)=T-T_{0}$, where $\textrm{T}_{0} = 57330.8237$ (MJD). }
    \label{fig:nustar_lc}
\end{figure}
We considered the \textit{NuSTAR} observation 30102041004 (start time $\textrm{t}_{\textrm{MJD}} = 57330.8237$, exposure: 21.88 ks), analyzed also in \cite{2018AstL...44..149S} and \cite{brumback2020}, taken when the source was close to the super-orbital peak ($\psi_{\mathrm{sup}} = 0.50-0.55$) and also out-of-eclipse respect to the orbital modulation (see \cite{2018AstL...44..149S} for details). We extracted the source light curve, shown in Fig. \ref{fig:nustar_lc} with a time bin of 200s, and spectra with the tools \textit{nupipeline} and \textit{nuproducts} for both the instruments FPMA and FPMB aboard the satellite using a circle region with a radius of $120''$ centered on the source position. We selected a background region of the same size in a portion of the image distant from the source. Spectral channels were re-binned following \cite{optimal_binning} so to not oversample the instrument energy resolution by more than a factor of three and to have at least 20 counts per bin in the energy range 3-50 keV.
\section{Data screening}
\label{sec:data screening}
To study the correlated soft and hard X-ray spectral behaviour of LMC X-4 even along a single super-orbital cycle requires a significant number of pointings. For the hard X-ray band, we can use and average the BAT survey data, but for the soft X-ray band no such  spectroscopic survey tool exists and we must rely on this single observational campaing to infer the general behaviour of the source.
However, if we compare the XRT and BAT lightcurves of the period that spans the XRT monitoring (Fig. \ref{fig:xrt_total_lcurve}), we notice that they both clearly follow the super-orbital 30.4 d modulation that has been persistently observed for the high energy data.
In the following we will assume that the randomly super-orbital cycle sampled with XRT and analyzed in this paper, is representative of the long-term behaviour of LMC X-4. 
Similarly to the BAT spectral extraction (see Sect. \ref{sub:bat_data}), we phase-selected the XRT data into 20 phase-bins of equal duration. To avoid the contamination of the effects of the orbital motion, we rejected the XRT intervals where the source was in eclipse. However, we recall that previous authors found that the flux variation induced by the super-orbital modulation is orders of magnitudes stronger than that induced by the orbital motion. Therefore we do not expect that orbital-dependent or local effects would considerably affect our results. We used the FTOOLS \textit{addspec} to sum the spectra and combine the response matrices for observations in the same orbital phase. We summarise the results from this screening, namely the association of the XRT ObsID and the phase-selected interval of the XRT data that will be analyzed in the remaining of this work, in Table \ref{tab:fasi_obs}. We are left with 15 out of the 20 super-orbital phases selected for our analysis. In particular, we will not consider for the broad-band analysis the following phases $\psi_{\text{sup}}$: 0.05-0.10, 0.15-0.20, 0.50-0.55, 0.75-0.80, 0.95-1.00.
\begin{table}
    \small{$^{a}$ Last two digits of the \textit{Swift}/XRT OBSID: 000335380XX}\\
    \centering
     \begin{tabular}{l |l }
    \hline
    $\Psi_{sup}$ & XRT ObsID$^{a}$ \\ 
    \hline
    $0.00-0.05 $ & 34 \\
    \hline
    $0.10-0.15$  &  38, 39 \\
    \hline
    $0.20-0.25$ & 12 \\
    \hline
    $ 0.25- 0.30$ & 13\\
    \hline
    $0.30-0.35 $ &14, 15\\
    \hline
    $0.35-0.40 $ & 16 \\
    \hline
    $0.40-0.45 $ & 17\\
    \hline
    $\mathbf{0.45-0.50} $ & \textbf{19, 20} \\
    \hline
    $ 0.55-0.60$ & 22, 23, 43, 44 \\
    \hline
    $0.60-0.65 $ & 45 \\
    \hline
    $0.65-0.70$ & 25,47  \\
    \hline
   $0.70-0.75 $ & 26, 48 \\
    \hline
    $0.80-0.85 $ & 29 \\
    \hline
    $0.85-0.90 $ & 30 \\
    \hline
    $0.90-0.95$ & 32\\
    \hline
    \end{tabular}
    \caption{List of the super-orbital phases (left) and correspondent XRT observations (right) included in the spectral analysis after the data screening described in Sec. \ref{sec:data screening}. The bold line refers to the high flux data through which we determined the physical properties of the X-ray spectrum. }\label{tab:fasi_obs}
\end{table}

\section{Search for super-orbital phase-resolved spectral changes}
\label{sec:pheno_superorbital}
In this section, we test simple models to the BAT and XRT super-orbital resolved spectra, independently, with the aim to describe the evolution of the spectral parameters along the super-orbital modulation. All the errors reported here and in the rest of the paper are at 90\%.
\subsection{15-60 keV energy range}
\label{sub:pheno_hard}
\begin{figure}
    \centering
    \includegraphics[width=\columnwidth]{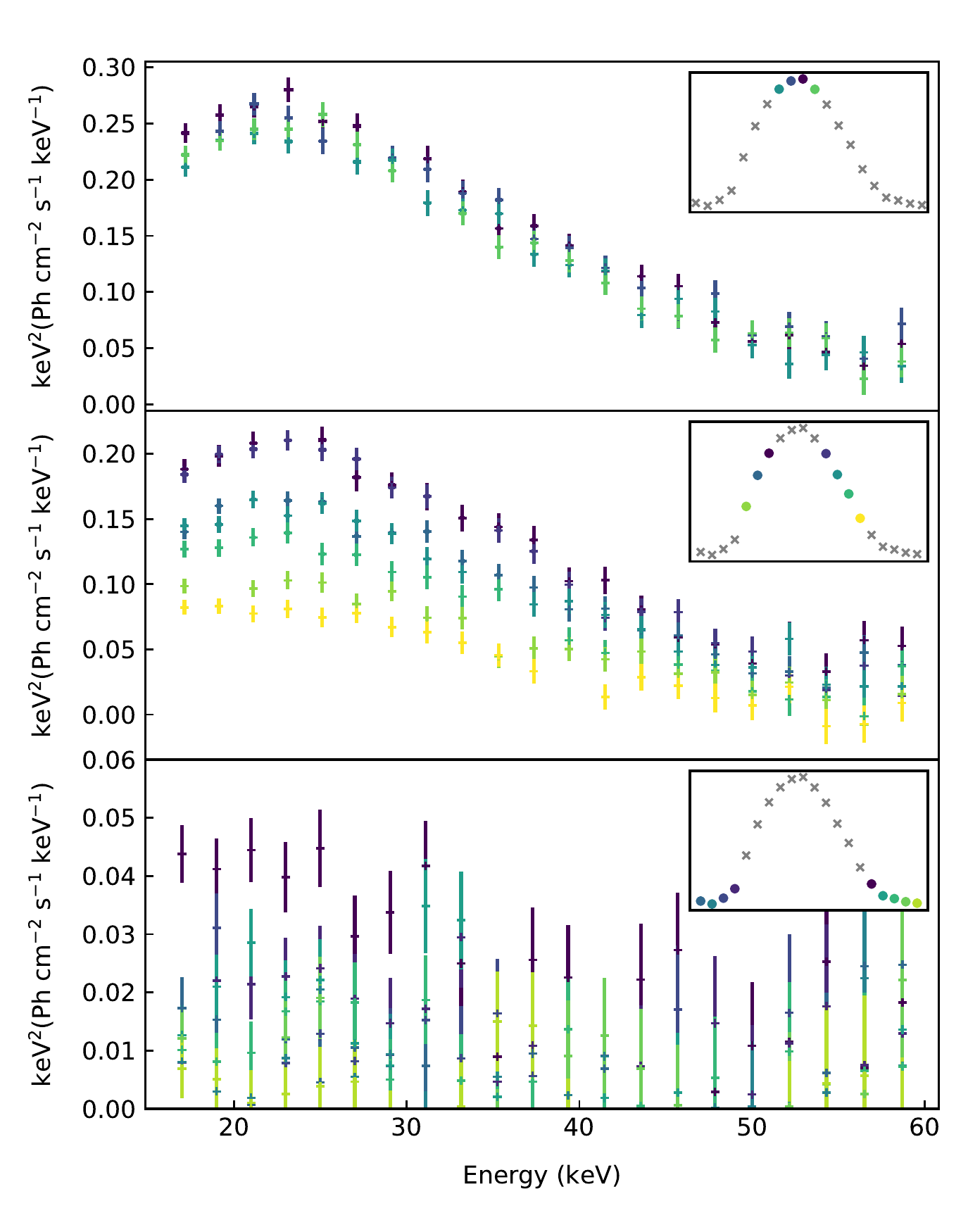}
    \caption{BAT super-orbital resolved spectra unfolded through a power-law model with photon index = 0. Spectra have been divided into three groups, depending on their flux level, for an easy visual inspection. The color code refers to the super-orbital phase to which the spectra belong, which is shown in the inset of each panel.}
    \label{fig:bat_folded_constant}
\end{figure}
\begin{figure}
    \centering
    \includegraphics[width=\columnwidth]{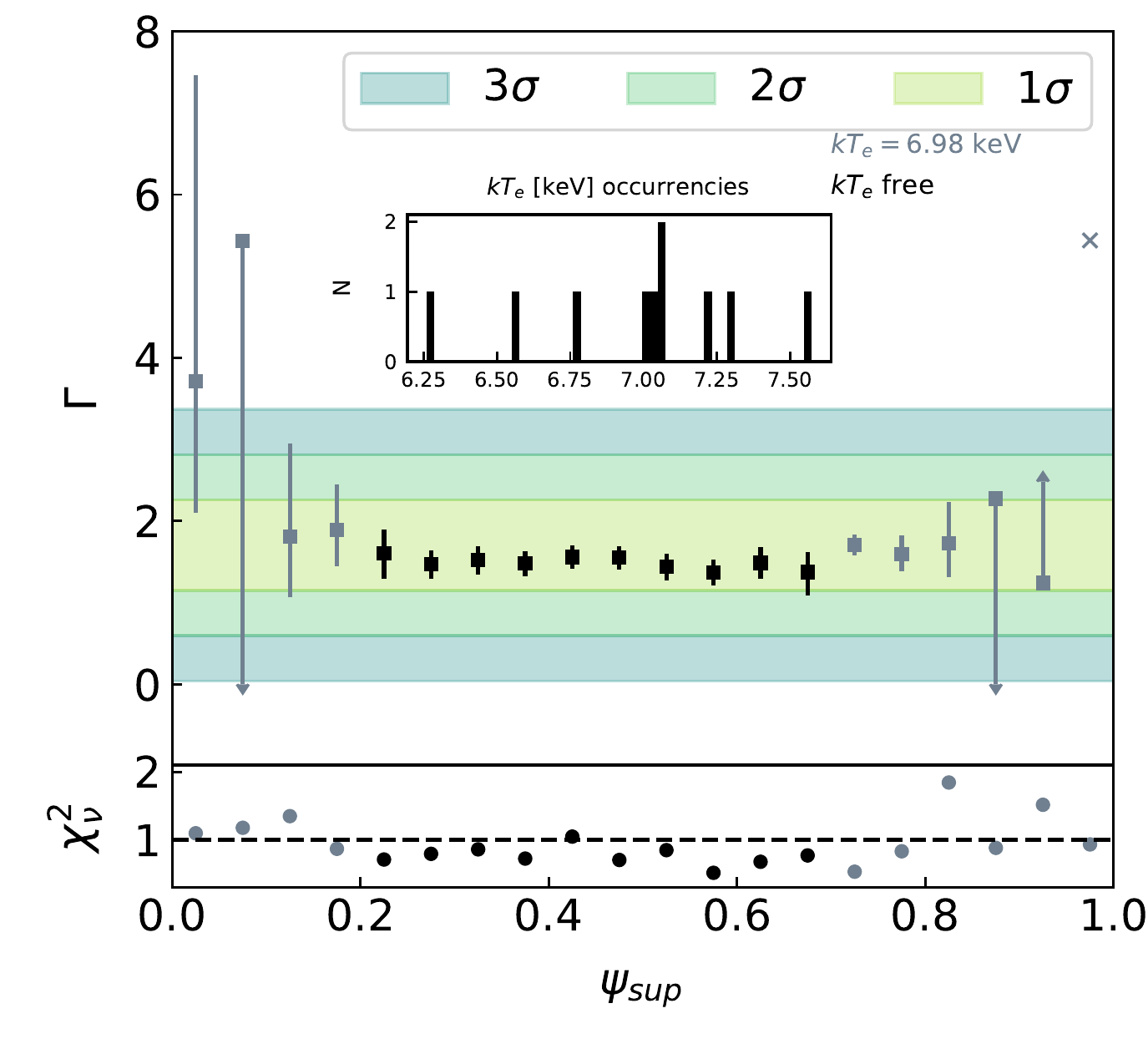}
    \caption{Fit of the BAT super-orbital phase-resolved spectra with the model \textsc{nthcomp}. \\ \textit{Upper panel.} Black points: all the parameters are left to vary. Inset: occurrencies of the determined values of $kT_{e}$ for the high flux phases (0.25-0.75). Grey points: fit results for the low flux spectra, with a fixed value of the electron temperature (k$T_{e}=6.98$ keV).  \textit{Lower panel:} reduced chi-square of each fit. }
    \label{fig:bat_pheno_superorb}
\end{figure}

In Fig.\ref{fig:bat_folded_constant} we show the phase-resolved BAT spectra unfolded through a power-law model with photon index = 0.\footnote{We applied systematics to the BAT spectra. Usually, systematics of the order of 2-3\%, depending on the statistics, on the background and on the position of the source in the field of view, are required.} We divided them into three groups, depending on their flux level, to make this plot easy to understand, because the flux level decreases considerably from the high state to the low state. From Fig. \ref{fig:bat_folded_constant} it is visually clear that the main difference consists in a gradual change in the flux level but not in the slope. We notice that the spectra which characterize the source around the minimum of the super-orbital phase (bottom panel) have a low signal-to-noise level.
 We then fitted all the spectra with the same model. 
We therefore used a simple Comptonization model (\textsc{nthcomp} in \textit{Xspec}) so that we retain a valid approximation for the accreting pulsar emission at this energy range. Initially all the model components have been left free to vary, except for the seed photon temperature ($kT_{bb}=0.1$ keV). We obtained good fits for the high flux spectra ($\psi_{\text{sup}}$ in range 0.25-0.75), while we found unconstrained values of the photon index, $\Gamma$, and the electron temperature, $kT_{e}$ for the phases with the lowest flux levels. This is not surprising. In fact, the majority of the energy channels in their 15-60 keV spectra consist of upper limits, preventing the determination of the spectral parameters. However, we noticed that the well constrained values of $\Gamma$ and $kT_{e}$ ($\psi_{sup}$ in range 0.25-0.75) are clustered around their correspondent mean values, 1.5 and 7 keV respectively within one standard deviation, leading us to conclude that our data do not require changes in the spectral parameters. We then tested whether this assumption can be extended also to the phases with lower counts, and fit them using the same model with the electron temperature fixed to $kT_{e} = 6.89$ keV, i.e. the mean value obtained for the high flux data. Results, presented in Fig. \ref{fig:bat_pheno_superorb}, show that 
the spectral photon index $\Gamma$, as well as its upper/lower limits, are consistent within $3\sigma$ along the different phases. The only spectrum with undetermined value of $\Gamma$ is the one with the lowest flux, which is labelled as an 'X' in Fig. \ref{fig:bat_pheno_superorb}. 

\subsection{0.3-10 keV energy range}
\label{sub:pheno_soft}
After having established the general  spectral quality of the hard X-ray data, we turned to the soft X-ray data from the Swift/XRT campaign and fitted the phase-selected XRT spectra with an absorbed phenomenological model which consists of a thermal component and a power-law (\textsc{tbabs*(diskbb+pow)} in \textit{Xspec}).
\begin{figure}
    \centering
    \includegraphics[width=\columnwidth]{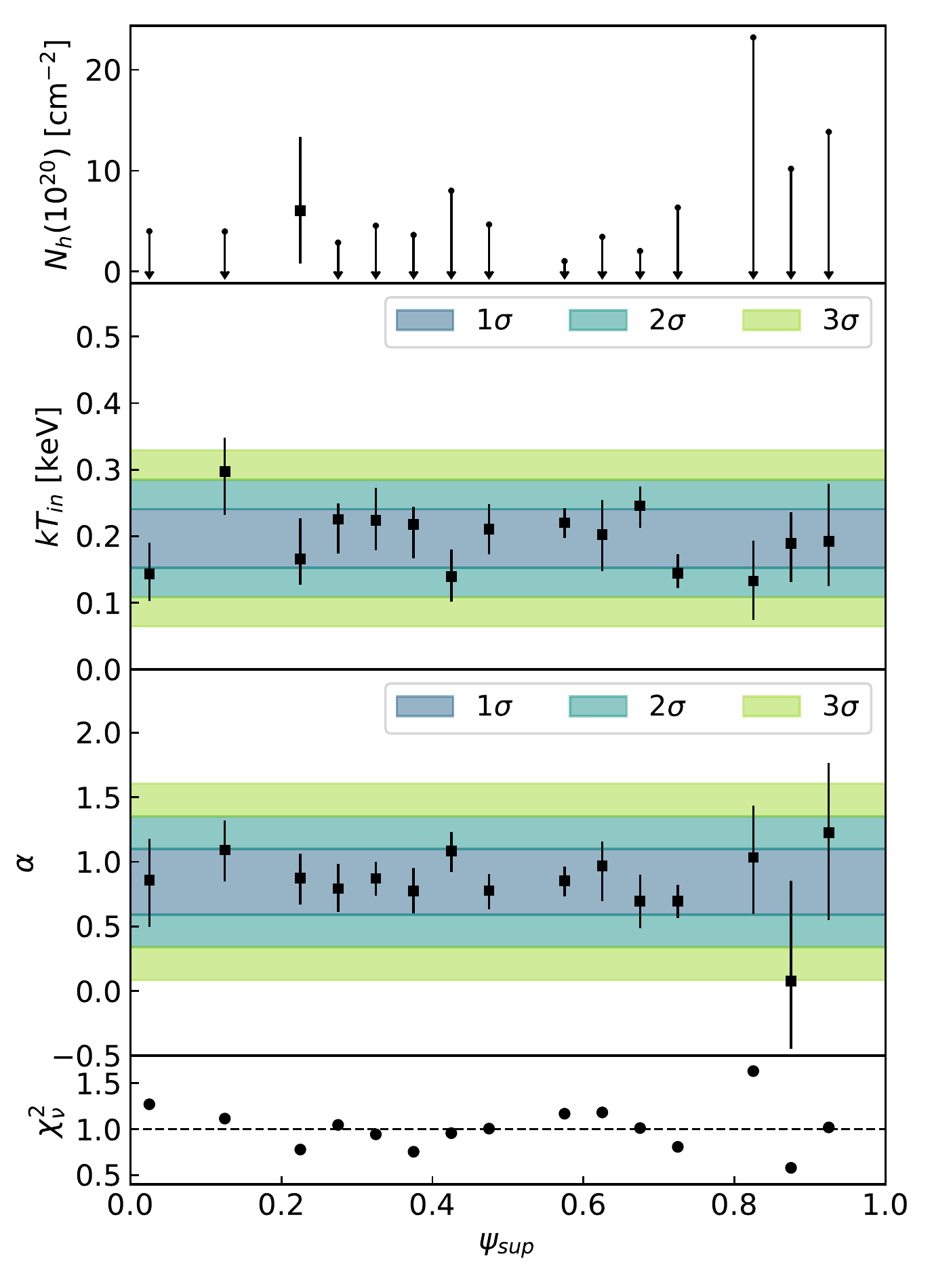}
    \caption{Fit of the XRT super-orbital phase-resolved spectra with the model \textsc{tbabs*(diskbb+pow)}.}
    \label{fig:xrt_pheno_superorb}
\end{figure}
The results obtained are displayed in Fig. \ref{fig:xrt_pheno_superorb}. The first panel shows the best fitting $N_{H}$ values for each spectrum. Given that the expected absorption value\footnote{Average value obtained through the HEASARC's N$_{\rm H}$ calculator: \url{https://heasarc.gsfc.nasa.gov/cgi-bin/Tools/w3nh/w3nh.pl})} is N$_{H}$ $ = 8\times10^{20}$cm$^{-2}$, we are not able to constrain it with our data. Moreover, this is an indication of no significant local absorption. This result is consistent with the findings of \cite{2003A&A...401..265N} with RXTE data. The second and third panels show instead that the model parameters are consistent  to one another at the different phases. We also note that the temperature of the soft component does not change along the super-orbital phases. All the values are consistent within one sigma as well as the photon index of the power-law.
\section{broad-band Spectral analysis of high flux epoch}
\label{sec:broad-band}
In this section we analyze the broad-band spectral properties of LMC X-4 at the flux peak of its superorbital period (phase 0.45-0.5). We jointly fit the \textit{NuSTAR} data and the XRT merged spectrum for the phase interval 0.45-0.50. Before doing that, we checked wether the joint fit of non simultanous data is feasible. First, a quick look at the \textit{NuSTAR} lightcurve shows that this observation is free of flares episodes, which have been observed to occur almost daily in this source (see Fig. \ref{fig:nustar_lc}), leading to an increase of the flux level, as well as changes of the spectral and timing properties (see \cite{2018AstL...44..149S,2018ApJ...861L...7B}). We first visually inspected the flux and spectral shape consistency of the XRT and NuSTAR datasets by over-plotting them in the same graph (Fig. \ref{fig:folded_jointfit} of Appendix \ref{app:useful_figures}).
%
In our modelling, we will use a multiplicative constant term, \textsc{const}, to take into account cross-calibration mismatches between the different instruments and flux variations due to the non simultaneity of the data. We fixed the one relative to the XRT data to unity and let free to vary those relative to the two \textit{NuSTAR} detectors (in the following we will refer to $C_{\text{NuA}}$ for the detector FPMA and to $C_{\text{NuB}}$ for the detector FPMB). All the uncertainties reported in this section are at 90$ \% $.\\

%
%
%
\subsection{Semi-phenomenological model}
\label{sec:pheno}
The broad-band X-ray spectrum of LMC X-4 is characterized by a very hard ($\Gamma < 1$) cut-off power-law (see e.g. \citealt{2003A&A...401..265N,2017AstL...43..175S,2021MNRAS.502.1163P} and references therein), plus a soft excess which is dominant below 1 keV and has usually been fitted with a black-body, a disk black-body, a bremsstrahlung component or a combination of them (\citealt{nielsen09_spec,2019ApJS..243...29A}). Moreover, this source exhibits a strong emission line at  $E \sim 6.5$ keV which can be fitted with a Gaussian of width $\sim 0.5$ keV (see e.g. \citealt{2017AstL...43..175S}).\\
\cite{coplrefl} interpreted the iron fluorescence line and the soft excess as due to a disk reflection component in which it is assumed that a constant-density accretion disk is irradiated by the cut-off power-law spectrum. We followed the interpretation of \cite{coplrefl} and used their model, \textsc{coplrefl} in \textsc{Xspec}, to jointly fit the soft excess and the emission line. 
In order to better constrain the \textsc{coplrefl}\footnote{\url{https://heasarc.gsfc.nasa.gov/xanadu/xspec/models/coplrefl.html}} parameters, it is essential to know the properties of the incident radiation, namely the photon index $\Gamma$, the cut-off energy $E_{\text{cut}}$ and the e-folding energy $E_{\text{fold}}$. We therefore fit the broad-band data with the model: \textsc{const*tbabs*(coplrefl+highecut*pow)}, where $\Gamma$, $E_{\text{cut}}$ and  $E_{\mathrm{fold}}$ of the reflection component are tied to those of the \textsc{highecut*pow} component. At first we let vary the absorption column density, but we found an upper limit ($ N_{\rm H} \leq 2 \times 10^{20}$ cm$^{-2}$) which is lower with respect to the known Galactic value towards LMC X-4\footnote{Average value obtained through the HEASARC's N$_{\rm H}$ calculator: \url{https://heasarc.gsfc.nasa.gov/cgi-bin/Tools/w3nh/w3nh.pl})} ($ N_{\rm H} = 8 \times 10^{20}$ cm$^{-2}$). We therefore decided to fix it to its Galactic value. 
Evident residuals below 0.6 keV and some at $\sim$ 1 keV were present. \cite{nielsen09_spec} analyzed the emission bump at $\sim 1$ keV with high resolution spectroscopy and found that, when the system is approaching the high state, this bump is characterized by narrow emission lines (helium-like triplets of nitrogen and oxygen and $Ly\alpha$ lines of nitrogen and oxygen). We were not able to resolve these lines with the \textit{Swift} spectra due to low energy resolution and statistics of the XRT spectra. 
%
\begin{figure*}
    \centering
    \includegraphics[scale=0.8]{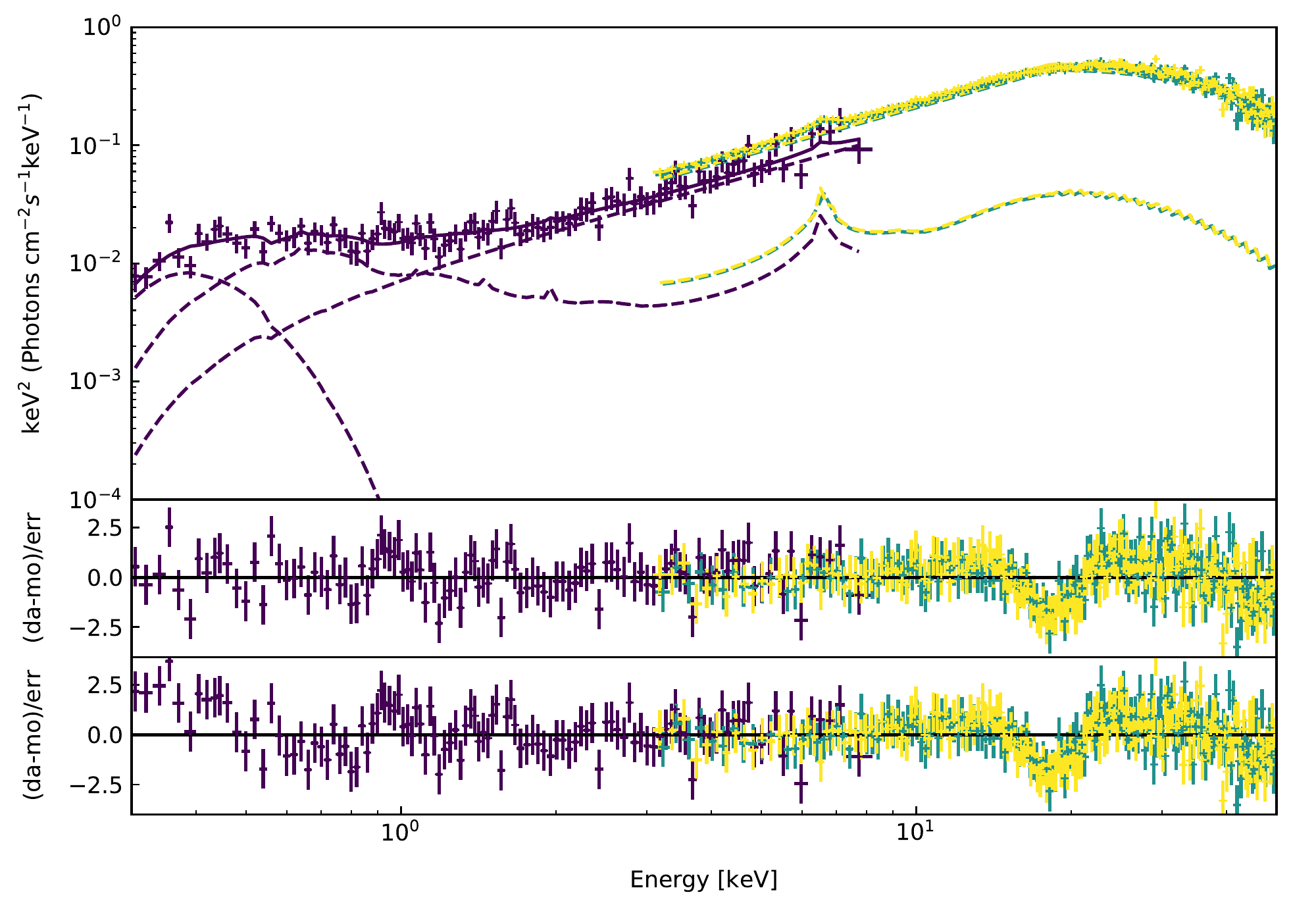}\caption{Best fit of the high flux data with the phenomenological model \textsc{const*TBabs*(diskbb+coplrefl+highecut*pow)} (upper panel) and its residuals (middle panel). Swift/XRT data, \textit{NuSTAR}A and \textit{NuSTAR}B are represented in violet, azure and yellow, respectively (from dark to lighter colors for a greyscale visualization). Lower panel: the residuals obtained without the soft black-body component provides evidence that an additional component is needed to flatten the soft excess at lower energies.}\label{fig:fit_pheno}
    \includegraphics[scale=0.8]{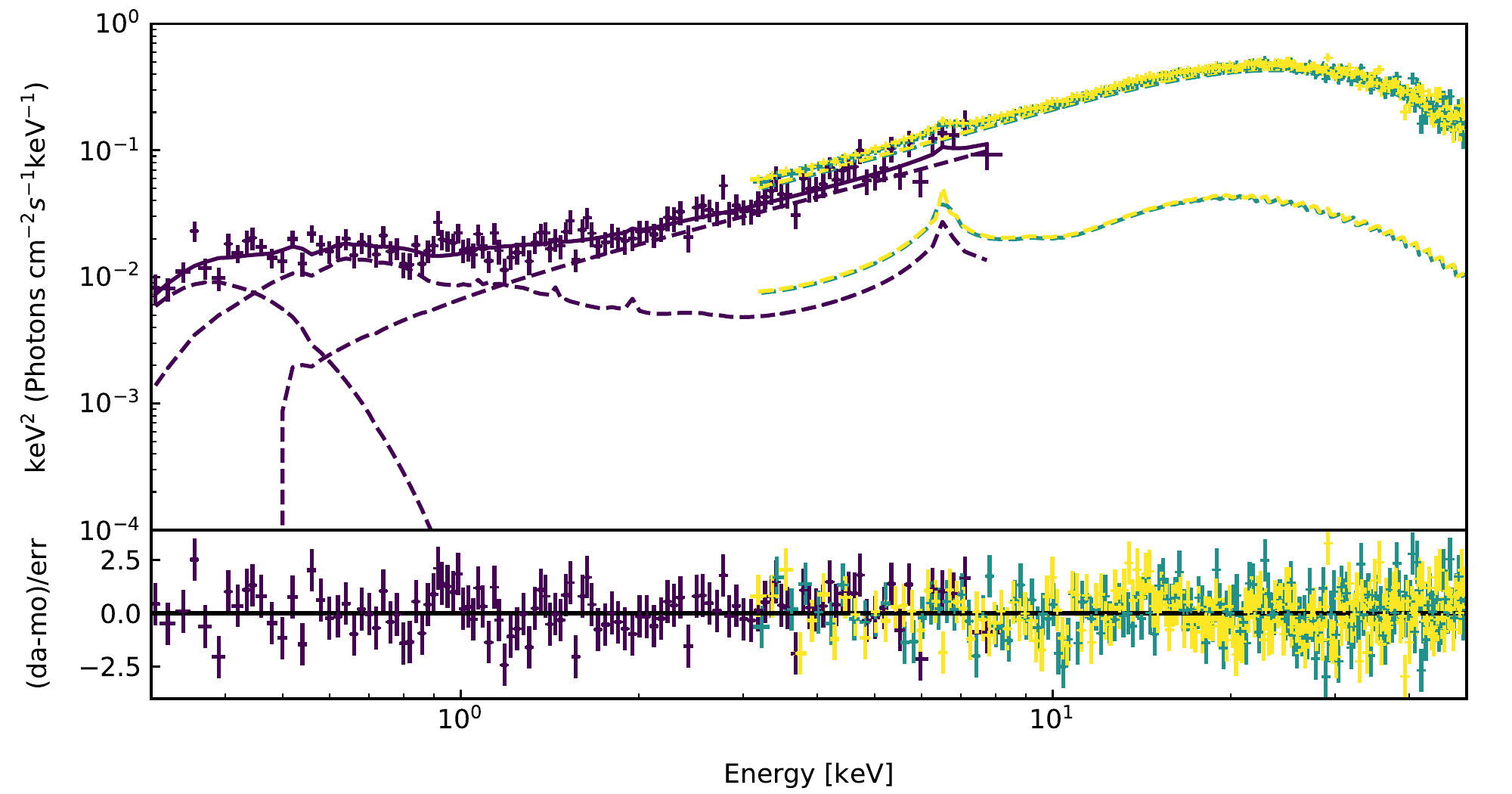}\caption{Best fit of the high flux data with the physical model \textsc{const*tbabs*(diskbb+coplrefl+bwcycl)} (upper panel) and residuals (lower panel).}\label{fig:best_bw}
\end{figure*}
The lower panel of Fig. \ref{fig:fit_pheno} shows that the residuals below 0.6 keV are significant. Therefore we added to our model a thermal component.  
The iron-like emission line, as well as the soft excess down to 0.6 keV are well fitted by the reflection of the illuminating hard X-ray flux, with $z \sim 0.035$ and ionization parameter $\log\xi \sim 3.06$. 
The reflection emission indicates the presence of an accretion disk, so that we added a multi-colour disk black-body. Our final model is: \textsc{const*tbabs*(diskbb+coplrefl+highecut*pow)}. 
The best fit with model components are shown in Fig. \ref{fig:fit_pheno} and all the parameters are listed in Tab. \ref{tab:bestfit} (left). 
%
%

\subsection{Physical model}
\label{sec:phys_model}
In this section we apply a physical model to the spectrum of LMC X-4, with the aim of characterizing the properties of its accretion flow. Matter accreting onto a strongly magnetized ($ B \sim 10^{12-13}\,\text{G}$) NS via a disk finally enters the magnetosphere, where the magnetic pressure overcomes the ram pressure. Here, the plasma couples with the field lines and is eventually channeled into a cylindrical accretion column (\citealt{1978ApJ...223L..83G,2007rapp.book.....G}). 
For systems with luminosity in the range $10^{36}-10^{38}$ \unilum, the in-falling plasma encounters a radiation-dominated radiative shock region above the stellar surface of the NS \citep{1976MNRAS.175..395B}, through which the kinetic energy of the gas is removed, so that it can settle on the NS surface. \\
The dynamics of the accreting gas into the radiation pressure dominated accretion column have been investigated by \cite{bw_paper1}, hereafter BW07, where an analytical solution for the emitted spectrum is shown. The radiation inside the column is injected by three sources: black-body radiation is produced by the thermal mound at the column base, while cyclotron and bremsstrahlung radiations are produced throughout the column. They model bremsstrahlung and cyclotron separately, but they are actually connected, see \cite{riffert1999}, but also some discussion in \cite{farinelli2012}.The processes of bulk and thermal Comptonization of these seed photons by the in-falling plasma takes place, so that the hard X-rays can escape through the column.
Comprehensive details of this model are found in BW07 and \cite{2016ApJ...831..194W}.\\ In the following, we fit our data with its \textsc{Xspec} implementation: \textsc{bwcycl} \citep{2009A&A...498..825F}. 
\begin{table*}
\begin{tabular}{l|l|l|l|l|l}
\hline
 \multicolumn{3}{c}{Phenomenological Model} & \multicolumn{3}{c}{Physical Model}\\
 \hline
           \textsc{tbabs} & $N_{\mathrm{H}}\,\, (10^{22} \text{cm}^{-2})$ & 0.08 (fixed) & \textsc{tbabs} & $N_{\mathrm{H}}\,\, (10^{22} \text{cm}^{-2})$ & 0.08 (fixed) \\
            \textsc{diskbb}& ${kT}_{\mathrm{in}}\, \text{(keV)}$  & $0.068_{-0.018}^{+0.025} $  & \textsc{diskbb}& $kT_{\mathrm{in}}\, \text{(keV)}$  & $0.065_{-0.014}^{+0.018}$\\
            &$\text{Norm}$ & $(5_{-4}^{+31})\times 10^{5}$& &$\text{Norm}$ &$(7_{-6}^{+45})\times10^{5}$ \\
             \textsc{coplrefl} & $\log (\xi) $ & $ 3.03\pm 0.05$  & \textsc{coplrefl} & $\log (\xi) $ & $ 3.04 \pm 0.04$\\
              & $\Gamma$ & $0.775 $ (linked)  &  &$\Gamma$ &0.775 (fixed) \\
              & $E_{\text{cut}} \text{(keV)}$ & 18.20 (linked)  & &$E_{\text{cut}} \text{(keV)}$ &18.20 (fixed) \\
            &$E_{\text{fold}} \text{(keV)}$  & 14.99 (linked) & & $E_{\text{fold}} \text{(keV)}$&14.99 (fixed) \\
            & \textit{z}   & $0.029_{-0.014}^{+0.015}$ &  & \textit{z} & $0.037\pm 0.007 $\\
            & Norm  & $(1.28\pm 0.21) \times 10^{-31}$&  & Norm&$(1.37 \pm  0.11) \times 10^{-31}$ \\
            \textsc{highecut*pow} & $\Gamma$ & $ 0.775 \pm 0.017 $&  \textsc{bwcycl} & $\xi$ &$2.9_{-0.4}^{+0.8}$ \\
             & $E_{\text{cut}}$ (keV) &$ 18.2\pm 0.4$ & & $\delta$  & $0.41 \pm 0.11$ \\
            & $E_{\mathrm{fold}}$& $15.0 \pm 0.3$& &kT$_{\mathrm{e}}\, \text{(keV)}$  & $6.20 \pm 0.17 $ \\
             & & & & $r_{0}\, \text{(m)}$ & $855_{-90}^{+120}$ \\
             $C_{\text{NuA}}$&  &$1.48 \pm 0.05$ & & & $1.53_{-0.08}^{+0.09}$ \\
             $C_{\text{NuB}}$& & $1.52 \pm 0.05$& & & $1.57_{-0.09}^{+0.10}$\\
             \hline
             \chired (d.o.f)  & & 1.09 (446) & & & 1.08 (446) \\
            \hline
\end{tabular}\caption{Fit results for the semi-phenomenological model (left) and the physical model (right) of the high flux data. $C_{\text{NuA}}$ and $C_{\text{NuB}}$ are the values obtained for the \textsc{const} term of the two \textit{NuSTAR} detectors, respectively.}\label{tab:bestfit}
\end{table*}
\subsubsection{Self-consistent physical model} \label{sec:best_phys}
\begin{figure}
    \centering
    \includegraphics[width=\columnwidth]{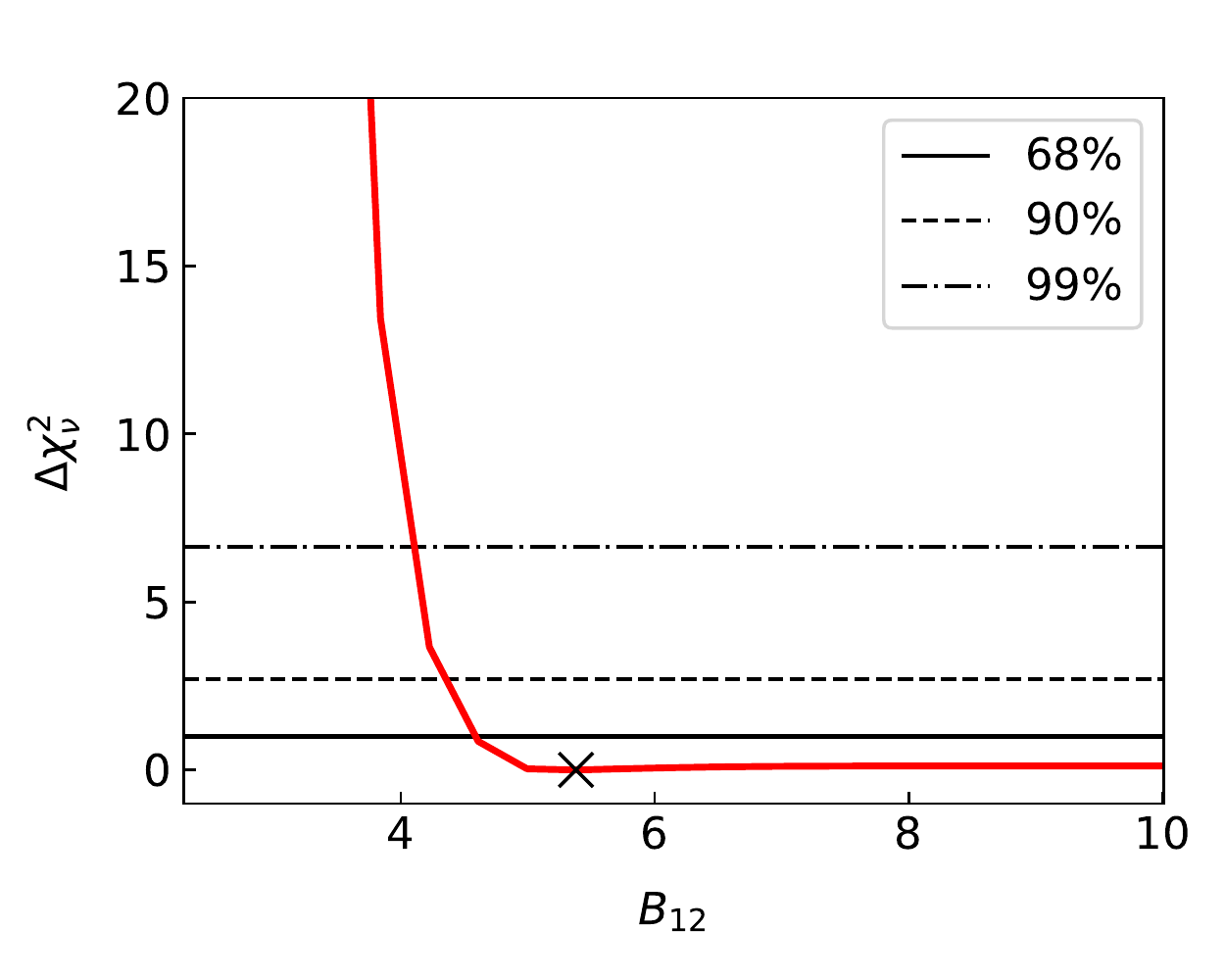} 
    \caption{Variation of the fit statistic ($\Delta$\chired) as function of the NS magnetic field in units of $10^{12}$G (B$_{12}$). }
    \label{fig:chired_bfield}
\end{figure}

Our physical model can be written in \textsc{Xspec} as  follows: \textsc{const*tbabs*(diskbb+coplrefl+bwcycl)}. The BW model requires some parameters to be fixed: the mass, the radius, the magnetic field strength of the NS and the mass accretion rate. We fixed the NS mass and radius at the canonical values (10 km and 1.4\footnote{We are aware that the actual NS mass is 1.57 \msol. However, the \textsc{bwcycl} model does not depend on it, so that we fixed it at its standard value.} \msol, respectively), as suggested in the guidelines of the model,\footnote{\url{https://heasarc.gsfc.nasa.gov/xanadu/xspec/manual/node148.html}} and constrained the mass transfer rate from the luminosity emitted within the \textsc{highecut*pow} component ($\dot{M}=1.2 \times 10^{18}$ g/s, assuming that all the gravitational energy of the material is converted into electromagnetic energy, i.e. efficiency = 1). We accounted for different values of the magnetic field strength, in the range $2.3\times 10^{12}$ - $1.0\times 10^{13}$ G. The disk black-body component was free to vary as well as the redshift, the ionization parameter and the normalization of the reflection component. $\Gamma$, $E_{\mathrm{cut}}$ and $E\mathrm{_{fold}}$ of the \textsc{coplrefl} component were fixed to the best-fitting values of the highecut*pow model discussed in Sect. \ref{sec:pheno}. We found that the fit was unacceptable for magnetic fields lower than $ 3 \times 10^{12}$ G ($1.54 \leq$ \chired $\leq 2.79$). We obtained the best fit for  $ B = 5.4 \times 10^{12}$ G (\chired = 1.08) and found that, for higher values of $B$, the fit becomes insensitive to it, therefore we fixed this value as input parameter for this model. This is shown in Fig.\ref{fig:chired_bfield}, where the variation of the $\Delta$\chist as function of the magnetic field is calculated for $B$ in the range $2.3\times 10^{12}$ - $1.0\times 10^{13}$ G in 20 linear steps.
We found a disk temperature of $kT_{\mathrm{in}}$ $\sim$ 65 eV and normalization of $\sim 7 \times 10^{5}$. 
The reflection component is still required and also in this case we found that it is produced in a region of high ionization ($\log \xi \sim 3.0$), with redshift  $z \sim$ 0.037.
Best-fit parameters are reported in Tab. \ref{tab:bestfit} (right). We notice that the reduced chi-squared values for the phenomenological and the physical model are very close (\chired 1.09 and 1.08, respectively). 
However, the pattern of residuals, which clearly shows the model discontinuity around 18 keV in the highecut model, is no longer noticed when the BW model is adopted.
\begin{table*}
    \centering
    \begin{tabular}{c|c|c|c|c|c|c}
    \hline
         $J (\text{g}\, \text{s}^{-1} \, \text{cm}^{-2})$& ${T}_{th}$ (K) & v$_{\mathrm{th}}/c$&$\text{z}_{\text{th}}$ (cm)&  z$_{\text{trap}}\, \text{(cm)}$ &z$_{\text{sp}}$ (cm) & L$_{\mathrm{crit}}$ (erg/s)\\
        \hline
          $5.25\times 10^{7} $& $2.04\times 10^{7}$& 0.02 & $1.09\times10^{4}$ & $2.10\times10^{7}$  &$ 3.57\times 10^{6}$ & $5.52\times10^{38}$\\
        \hline
    \end{tabular}
    \caption{Characteristics of the accretion column and the accreted matter obtained from the best fit model parameters. The related formulas are listed in Appendix \ref{app:Characateristics of the accretion column}}
    \label{tab:bw_phys}
\end{table*}
%
\subsubsection{The thermal component}
\label{sub:thermal}
%
%
%
From the \textsc{diskbb} normalization parameter we derived the inner disk radius, $R_{\mathrm{in}}$ $\sim$\, 4.1$\times$10$^{3}$ km (for an inclination $\theta \sim 59$\textdegree). For accretion to take place, $R_{\mathrm{in}} \gtrsim R_{\mathrm{A}}$, where $R_{\mathrm{A}}$ is the Alfvèn radius. Given our estimated lower limit on the NS magnetic field, we can set a lower limit on $R_{\mathrm{A}}$, according to the classical formula (see e.g. eq. 111 in BW07):
\begin{equation}
    R_{A} = 2.6 \times 10^{8} \bigg(\frac{B}{10^{12} \text{G}}\bigg)^{4/7}\bigg(\frac{R_{*}}{10 \text{km}} \bigg)^{10/7}\bigg(\frac{M_{*}}{M_{\odot}} \bigg)^{1/7} \bigg(\frac{ L_{X}}{10^{37} \text{\unilum}} \bigg)^{-2/7}
\end{equation}
where $B$, $R_{*}$, $M_{*}$ and $L_{\mathrm{X}}$ are respectively the magnetic field, the radius and the mass of the NS and L$_{X}$ is the accretion luminosity. From the result obtained with the best fit model ($L_{X}= 2.15\times 10^{38}$ \unilum, $B=6.0\times10^{12}$ G), we estimated $R_{\mathrm{A}} \sim $ 3.1$\times 10^{3}$ km, so that the condition $R_{\mathrm{in}} \gtrsim R_{\mathrm{A}}$ holds. 

\subsubsection{Properties of the accretion flow}
Once we have fitted the broad-band spectrum between 0.3-50 keV of LMC X-4 with a self-consistent physical model, we are able to shed light on the physical properties of the accretion flow. \\
We retrieved the properties of the accretion flow with the best fit values (see Tab. \ref{tab:bestfit}) and some key parameters (see BW07) that result from them (Tab. \ref{tab:bw_phys}). We briefly list the corresponding equations in Appendix \ref{app:Characateristics of the accretion column}.
The continuum, in the 2-50 keV energy range, is strongly dominated by the emission and scattering processes which take place in the accretion column. The plasma at the magnetospheric radius decouples from the accretion disk and is channeled by the magnetic field lines into the accretion column with a typical free-fall velocity $v_{\text{ff}} \equiv \sqrt{2GM_{*}/R_{*}} \sim 0.3c$. From the spectral parameters obtained with the fitting procedure, we found that the accretion column has a radius $ r_{0} <$ 1 km. The plasma enters the column and encounters a radiation-dominated shock at the sonic point, located at $z_{\mathrm{sp}} \sim 3.6\times10^{6}$ cm (see Eq. \ref{eq:zsp}) from the surface of the NS. Inside the column, the processes of bulk and thermal Comptonization act in removing kinetic energy from the plasma, which finally settles to the surface of the NS, at rest, passing through the surface of the thermal mound with $v_{\mathrm{th}} = 0.02c$ and with mass accretion flux $J \sim 5.2\times10^{7}$ (g$\text{s}^{-1}\text{cm}^{-2}$). The thermal mound is just above the NS surface at height $z_{\mathrm{th}} \sim 1.1\times10^{4}$ cm with $ \mathrm{T} \sim 2\times10^{7}$ K. 
We found that inside the column the thermal Comptonization and the bulk processes are equally important in the energy balance between the electrons and photons populations: $\delta=4 y_{\text{bulk}}/y_{\text{th}} \sim 0.41 \sim 1$. Despite that, the presence of the cut-off energy of the spectrum  ($E_{\text{cut}} \sim 18$ keV, see Tab \ref{tab:bestfit} and Figs. \ref{fig:fit_pheno} and \ref{fig:best_bw}) is a clear signature of the strong contribution of thermal Comptonization in removing the high-energy photons. We found also that the photon trapping inside the accretion column takes place at $z_{\text{trap}} = 2.10\times10^{7}$ cm $\gtrsim z_{\text{sp}}$ (see Tab.\ref{tab:bw_phys}), so that photons are not able to escape so far beyond the sonic point. 
Reflection of the hard X-rays by an inner region of the accretion disk accounts for the fluorescent emission line observed at E $\sim 6.5$ keV and for the soft excess emission down to 0.6 keV (Fig. \ref{fig:best_bw}). 
The reflection spectrum is produced by a highly ionized material ($\log\xi \sim 3$), which is redshifted to $z \sim$0.037.
The high ionization of the reflecting material suggests that it is located in the inner parts of the accretion disk. \\
Finally, the softer X-ray emission of the spectrum, at  $E\leq 0.6$ keV, is dominated by the emission of the cold ($kT_{\rm in} = 0.06$ keV) standard accretion disk. 

\section{Super-orbital phase-resolved broad-band spectral analysis}
After having established the best-fit model for the brightest super-orbital phase and shown that the BAT and XRT spectra, taken separately, do not undergo substantial spectral changes, we analyzed how the joint hard and soft spectra varied along the other super-orbital phases using the best fit physical model.
We performed super-orbital phase-resolved broad-band spectroscopy with the XRT data collected during our monitoring campaign and the BAT data, selected in super-orbital phase.
\subsection{The high-energy modulation}
\label{sub:onlybat_fit}
\begin{figure}
    \centering
    \includegraphics[width=\columnwidth]{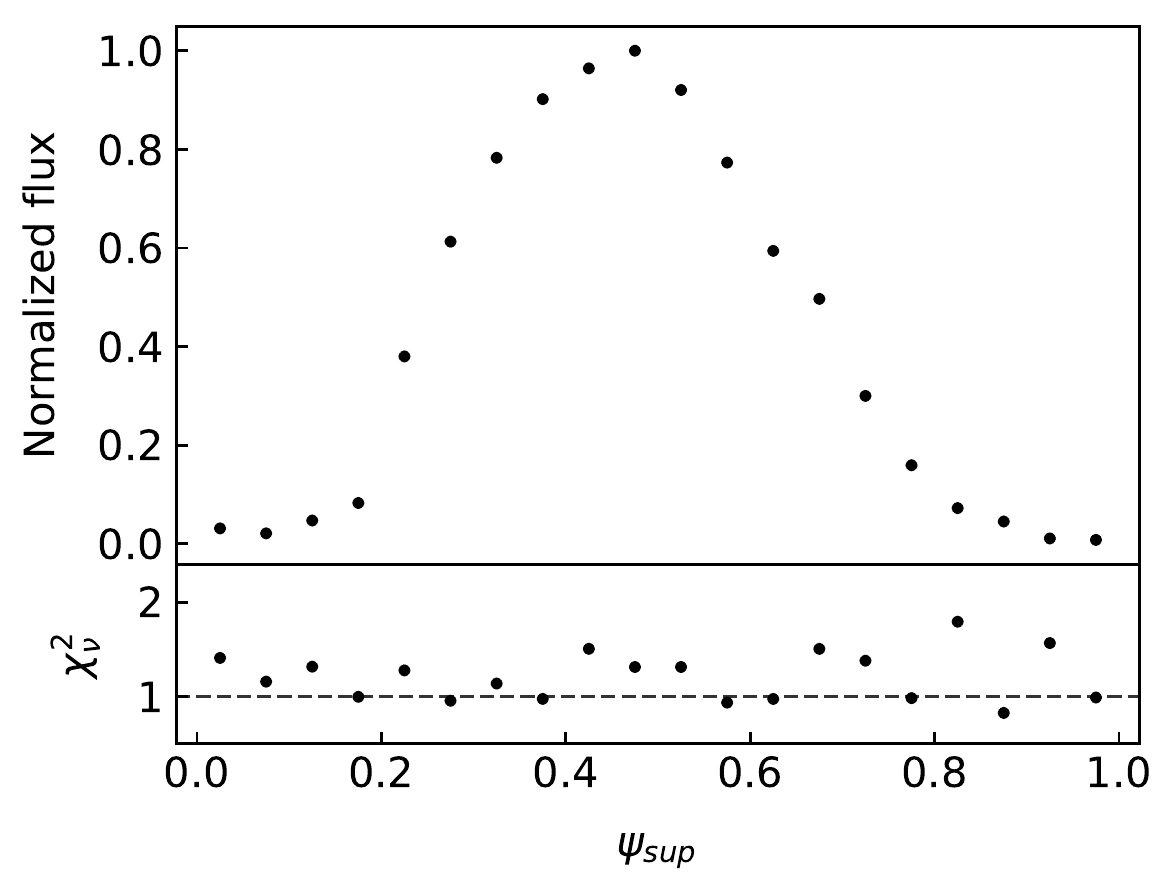}
    \caption{Upper panel: flux variation along the different super-orbital phases with respect to the maximum. Lower panel: \chired resulting from the fit of the super-orbital phase-resolved BAT spectra with the model \textsc{k$_{1}$*(coplrefl+bw)}, where all the model parameters are fixed, while the \textsc{k$_{1}$} term is the only free parameter.}
    \label{fig:onlyBAT}
\end{figure}

The hard X-ray emission of LMC X-4 has been analyzed by previous authors (see e. g. \citealt{1977ApJ...216..103E,2001ApJ...553..375L,2003A&A...401..265N,2005AstL...31..380T,brumback2020} and reference therein). So far, there is no evidence of spectral variability in the hard ($\gtrsim 20 $ keV) X-ray emission along the super-orbital period in agreement with our BAT phase-resolved analysis (Sec. \ref{sub:pheno_hard}). Given the shape consistency, we then fitted the data with the model \textsc{k$_{1}$*(coplrefl*bw)}\footnote{We did not consider here the absorbed \textsc{diskbb} component, which does not contribute to the emission at the BAT energy range.} where the multiplicative constant, \textsc{k$_{1}$},  
accounts for the fractional, phase-dependent, flux modulation. This is the only free parameter in each fit. 
In Fig. \ref{fig:onlyBAT} we show the flux variation (upper panel) and the \chired of the fits (lower panel) versus the super-orbital phase. We obtained acceptable fits, $ 0.79 \leq$ \chired $\leq 1.7$ (d.o.f. = 20), and we found \chired > 1.5 only for those low-flux state spectra, shown in the lower panel of Fig. \ref{fig:bat_folded_constant}, which have some high residuals in channels that are likely background dominated. 
This is a known issue for data taken with coded masks.

\subsection{Broad-band} \label{sec:broad_band_phresolved}
With the aim to further investigate on the super-orbital modulation of LMC X-4,
we performed a joint fit of the XRT and BAT spectra for each selected super-orbital phase (see Tab \ref{tab:fasi_obs}). 
We first considered the hypothesis in which the innermost regions around the central object, i.e. the accretion column and the disk, are uniformly shielded from the observer by an outer precessing ring. In this case, we would expect to observe a global (both soft and hard X-ray) flux variation along the different $\mathbf{\psi}_{\text{sup}}$ phases, that is a scaling of the model which describes the high flux data.\\
We used the model \textsc{const*tbabs*k$_{2}$*(diskbb+coplrefl+bw)}. Here, \textsc{const} is the cross-calibration coefficient (fixed to one for the XRT data and left free for the BAT data) and \textsc{k$_{2}$} represents the fractional change of the overall observed flux  with respect to the high flux data. The term \textsc{k$_{2}$} is a free parameter.
The spectral parameters of \textsc{diskbb}, \textsc{coplrefl} and \textsc{bwcycl} were fixed to the values obtained for the \textit{NuSTAR} and XRT joint fit (see Tab. \ref{tab:bestfit}, right) considering that we already excluded any spectral variation in the previous section. We fixed also their normalizations to the values obtained at the epoch of maximum flux, because our aim here is to test whether the flux variation is energy-independent
.
%
%
%
In this case, the coefficients \textsc{k$_{2}$} should be compatible with those derived from the high-energy data fits, \textsc{k}$_{1}$, representing the fraction of the emitted flux which is shielded to the observer.
\begin{figure}
    \centering
    \includegraphics[width=\columnwidth]{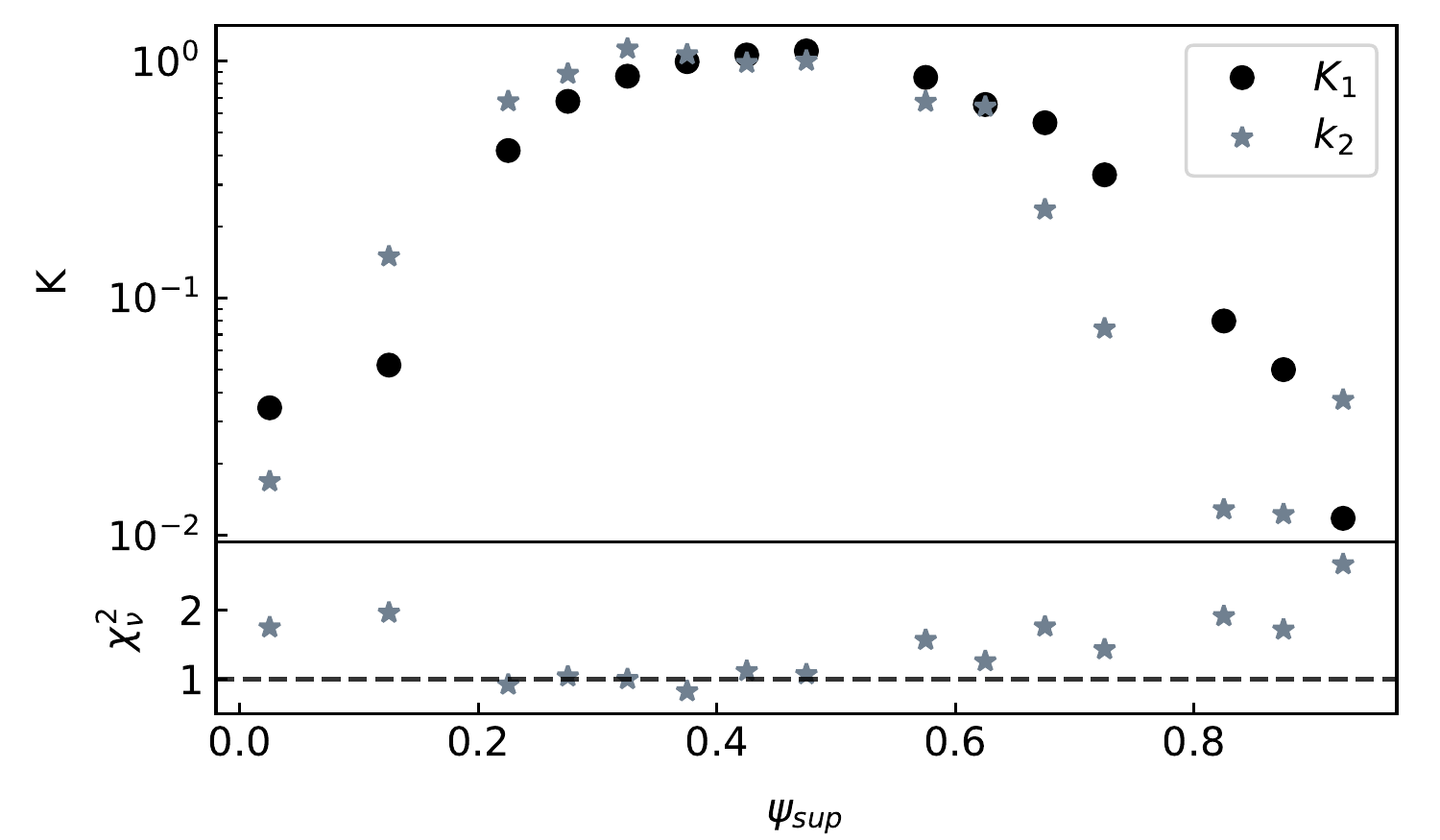}
    \caption{Upper panel: comparison of the coefficients \textsc{k$_{1}$} (black points) and \textsc{k$_{2}$} (grey stars); the magnitudes of the errorbars are contained in the size of the data points. Lower panel: \chired resulting from  the fit of the super-orbital phase-resolved broad-band (XRT+BAT) spectra with the model \textsc{const*tbabs*k$_{2}$*(diskbb+coplrefl+bw)}, where all the model parameters are fixed, while the \textsc{k$_{2}$} term and the \textsc{const} term for the BAT data are the only free parameters).} 
    \label{fig:test1}
\end{figure}
\begin{figure*}
\includegraphics[width=\textwidth]{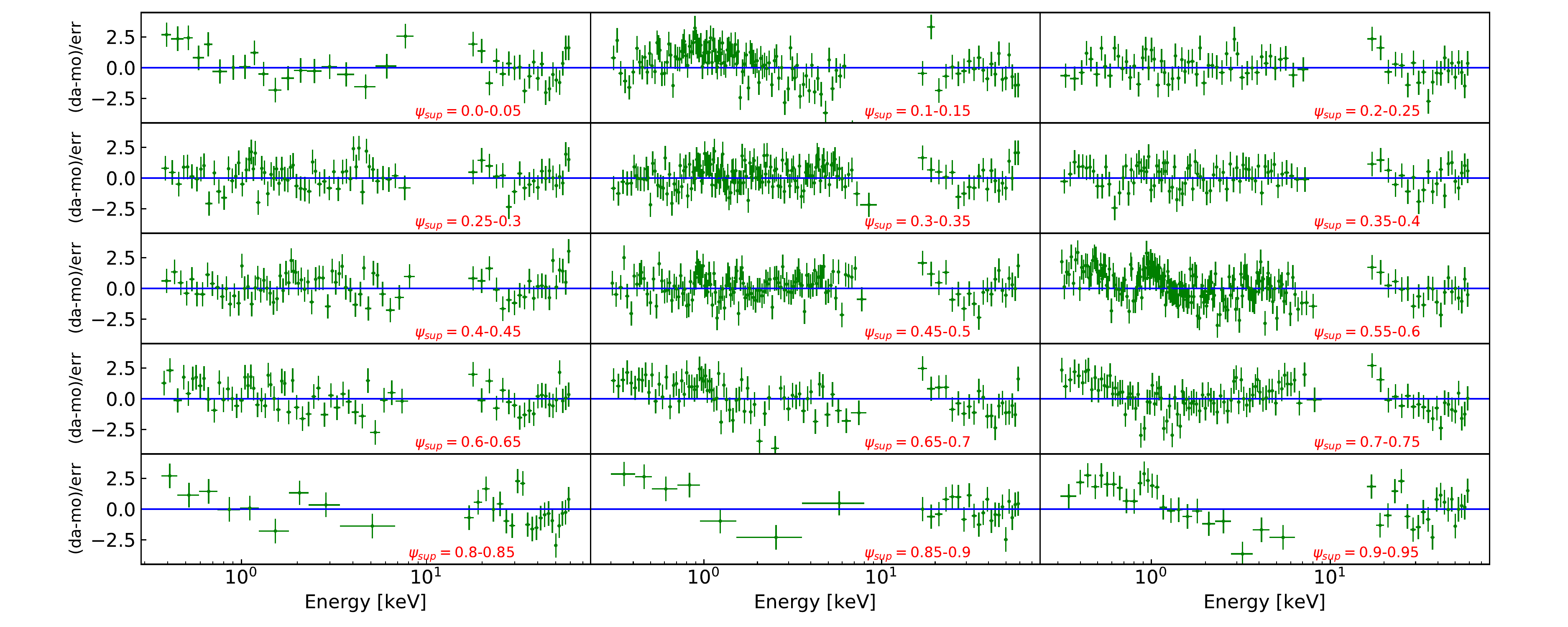}
    \caption{Residuals, for each XRT+BAT super-orbital phase-resolved spectra, of the fit with the model \textsc{const*tbabs*k$_{2}$*(diskbb+coplrefl+bw).}}
    \label{fig:res_Kmaxmodel}
\end{figure*}
\begin{figure*}
\includegraphics[width=\textwidth]{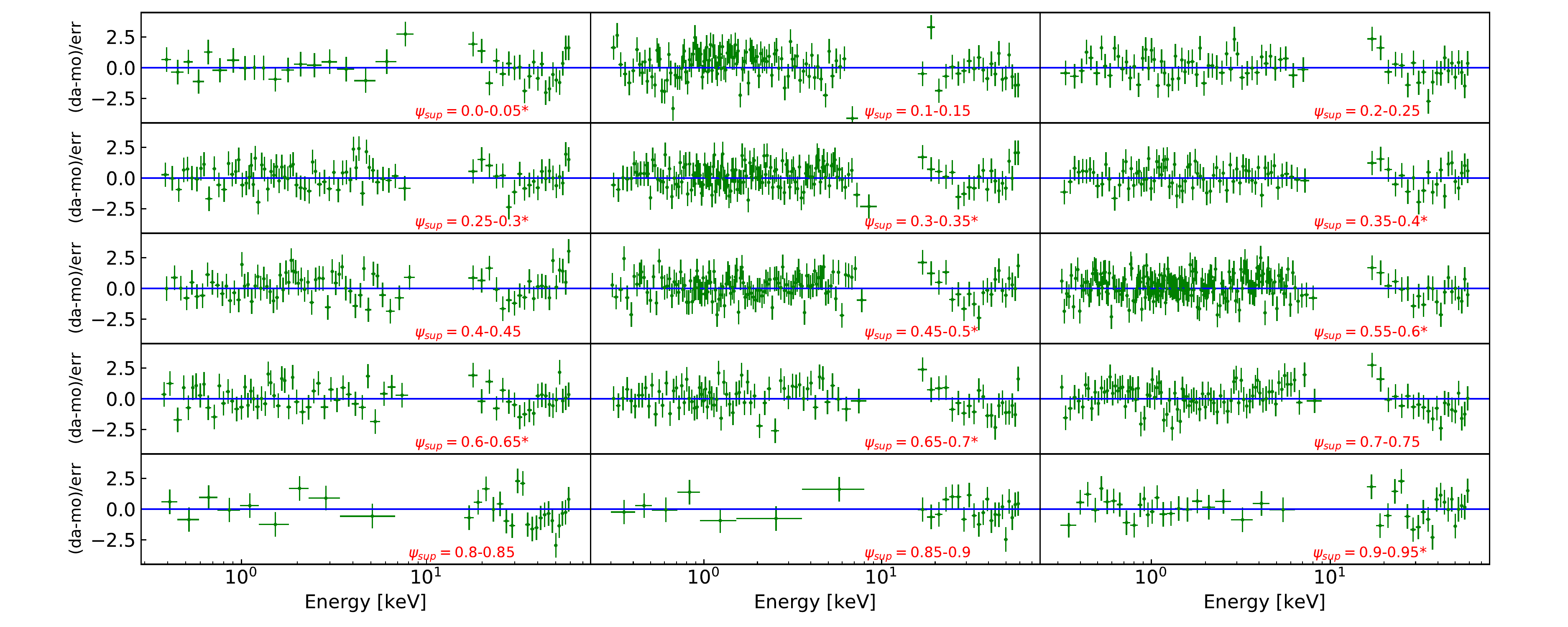}
    \caption{Residuals, for each XRT+BAT super-orbital phase-resolved spectra, of the fit with the model \textsc{const*tbabs*(diskbb+coplrefl+k$_{1}$*bw).} We marked with an asterisc (*) those phases for which we added an additional \textsc{gaussian} component at $\sim 1$ keV to flatten the residuals and improve the statistic (see text for details).}
    \label{fig:res_freenorms}
\end{figure*}
%
%
Our findings do not
support this assumption. In fact, the coefficients \textsc{k$_{2}$}, obtained for the broad-band data (grey stars on Fig. \ref{fig:test1}) are not consistent with those resulted only for high-energy data ( i.e. \textsc{k}$_{1}$, black points on Fig. \ref{fig:test1}). Moreover, the resultant statistic of these fits was not satisfactory ($0.83 \leq $\chired $\leq 2.67$). We found acceptable agreement, as expected, only for $\mathrm{\psi}_{ \mathrm{sup}}$ in the range 0.40-0.55. In Fig. \ref{fig:res_Kmaxmodel} we show the residuals obtained for each fit.\\
After having excluded that the entire accretion flow is uniformly shielded from the observer by an outer obscuring region, we proceeded further and searched for independent variation of the various components.
So we finally tested the broad-band data using the model \textsc{const*tbabs*(diskbb+coplrefl+\textsc{k$_{1}$}*bwcycl)}, where the normalizations of the \textsc{diskbb}, of \textsc{coplrefl} and the multiplicative \textsc{k$_{1}$} term are free to vary, while the cross-calibration term (\textsc{const}) has been fixed to unity only for the XRT spectra. The residuals obtained for each fit are shown in Fig. \ref{fig:res_freenorms}, where we marked with an asterisk (*) those phases for which an additional gaussian component was needed to represent the emission feature at $\sim 1$ keV previously analyzed by \cite{nielsen09_spec}.\footnote{The characterization of this component is outside the scope of this work and beyond the possibility of the low spectral resolution of the XRT data. However, the values we found for the line energy and width are consistent with those reported by \citealt{nielsen09_spec}.}
%
\begin{figure}
    \centering
    \includegraphics[width=\columnwidth]{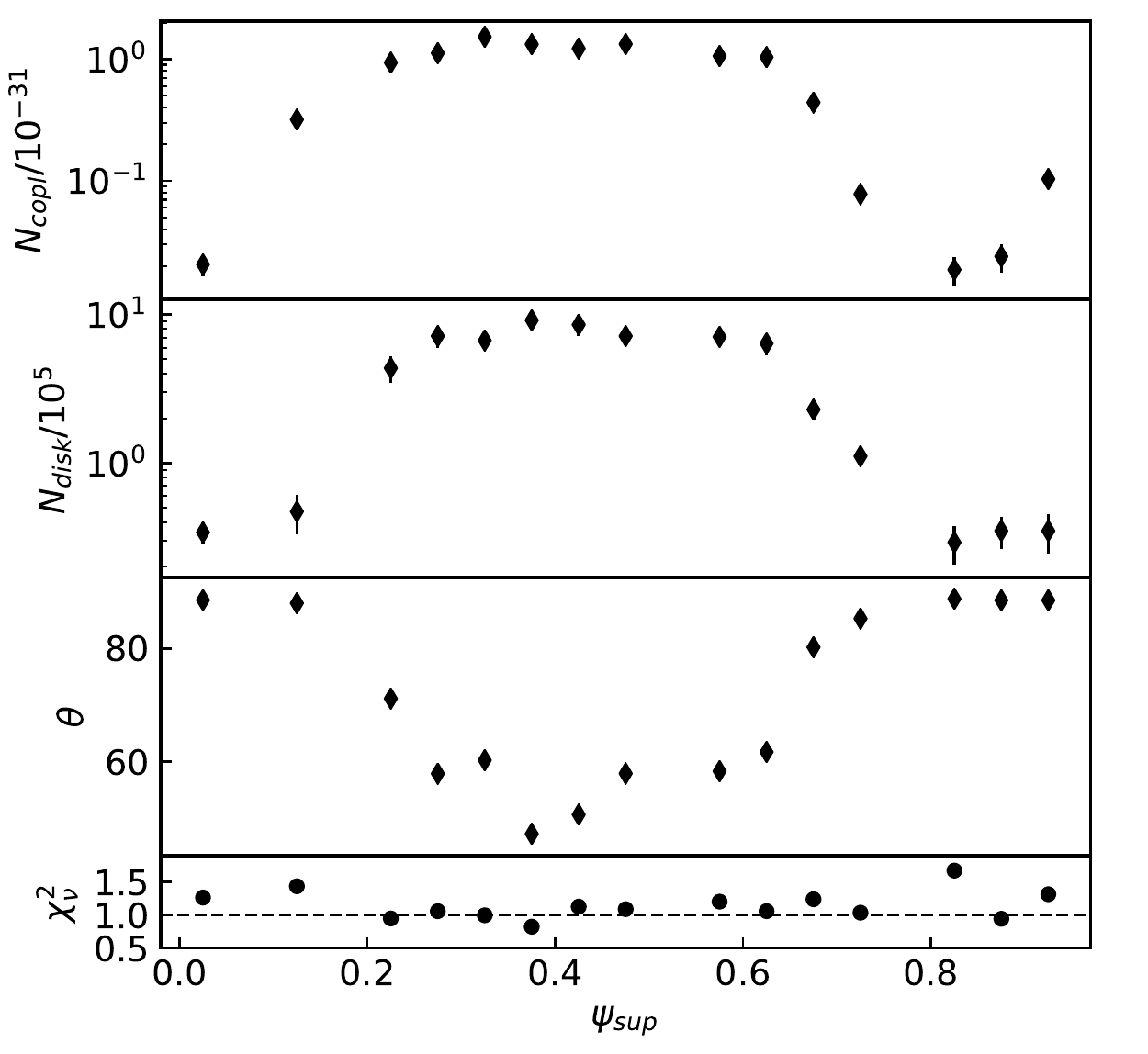}
    \caption{Variation of the model parameters for the fits obtained with the model \textsc{const*tbabs*(diskbb+coplrefl+k$_{1}$*bwcycl)}. We show how the normalization of the disk and the reflection components vary along the super-orbital phase, as well as the computed inclination of the outer tilted precessing disk (see text). In the lower panel we show the results of the fit statistic.}
    \label{fig:test_final_yeah}
\end{figure}
Fig. \ref{fig:res_Kmaxmodel} and Fig. \ref{fig:res_freenorms} show that we obtain bad fits if we scale the best fit model of the epoch of maximum flux with the same coefficient. On the contrary, the broad-band data are well fitted if we let free to vary the normalizations of all the model components.
Moreover, given that the fits are performed with all the model parameters (excluding the normalizations) frozen, this suggests that the disk temperature remains unchanged, as well as the physical properties of the plasma which interacts with the high-energy emission. Fig. \ref{fig:test_final_yeah} shows how the reflection and the disk normalizations (first and second panels) vary along the super-orbital phase. The change in the normalization of the \textsc{diskbb} component can be read as a statistically significant variation of the disk inclination, to which the normalization is linked via the following relation: $N_{\mathrm{disk}}=(R_{\mathrm{in}}/D_{10})^{2} \cos{\theta}$. Here, $\mathrm{D}_{10}$ is the distance normalized to 10 kpc, $\theta$  the disk inclination and the inner disk radius, $R_{\mathrm{in}}$, is fixed at the value found for the epoch of maximum flux (see Sec. \ref{sub:thermal}). The third panel of Fig. \ref{fig:test_final_yeah} shows how the disk inclination varies 
as a function of the super-orbital phase. If we consider that at $\psi_{\mathrm{sup}}$ the disc inclination is $\theta \approx 60$\textdegree, the amplitude of $\Delta \theta$ suggests that at the epoch of minimum flux the source is seen nearly edge-on, so that the innermost regions of the system are obscured to the observer. 

\section{discussion}
\label{sec:discussion}
We presented the first broad-band super-orbital phase-resolved spectral analysis of LMC X-4 resolved in 20 phase-bins. 
Our final purpose in doing this was to study the broad-band spectral variation along the super-orbital modulation. First, we constrained the broadband X-ray shape using a $NuSTAR$ observation with good statistics and an XRT observation in the same super-orbital phase adopting a self-consistent physical model. In this way, we could disentangle the contribution of the different processes to the overall emission and characterize the physical properties of the source. Then, once established the energy range in which each process dominates over the others, we studied their variation as functions of the super-orbital phase. 
\subsubsection*{Properties of the broad-band X-ray emission and characteristics of the accretion flow}
We used the data which correspond to the maximum of the super-orbital period ($\psi_{\mathrm{sup}} = 0.45-0.50$) to firmly characterize the properties of the accretion flow. The high-energy emission is produced mainly in the innermost regions, where the accreting plasma is channeled by the  magnetic field lines onto the accretion column, and then it settles on the surface of the neutron star. The emission produced within the accretion column strongly dominates the spectrum from $\sim$1.5 keV upwards. We modelled the soft excess as produced by two different processes: above $\approx$ 0.6 keV the major contribution comes from the disk reflection component, below 0.6 keV a colder disk component is dominant.
\subsubsection*{Properties of the accretion column}
We derived the properties of the accretion column adopting the \textsc{bwcycl} model. 
This is, to our knowledge, the first fit of an X-ray spectrum using this model. BW07 showed an application of it, without a proper fit by using broadband data from a $Beppo-SAX$ observation. They assumed a different set of input parameters: a NS magnetic field of $\sim$\,3 $\times$ 10$^{12}$ G, a distance of 55 kpc and a mass accretion  rate of 2\,$\times$\,10$^{18}$ g s$^{-1}$. To obtain a good match with the dataset, they also fixed the $\delta$, $\xi$, and $kT_e$ and $r_o$ parameters to 1.3, 1.15, 5.9$\times$10$^{7}$ K and 680 m,  respectively. Our results, on the contrary, are directly found through a broadband fitting of high-quality X-ray data.
Our data constrained a lower limit for the magnetic field of the neutron star of $B_{\mathrm{low}} \approx 6.0\times10^{12}$ G, which is in agreement with the lack of detection of any cyclotron absorption feature in the high SNR part of the X-ray spectrum and with the marginal detection found by \cite{2001ApJ...553..375L}  at $\approx$ 100 keV, which would correspond to $B\sim$ 10$^{13}$ G. Among the known young NS hosted in high-mass X-ray systems, such value would be not exceptionally higher than the average B-field value \citep[][]{2019A&A...622A..61S}, and might further indicate that the accretion process has not significantly lowered its native $B$ value \citep{2001ApJ...557..958C}.
%
In accordance with BW07, we found that the emitted spectrum is characterized by a strong contribution from thermal Comptonization, which causes the observed cut-off and the flattening of the spectrum at lower energies. 
\subsubsection*{Properties of the reflection component}
We interpreted the emission line at $\sim 6.5$ keV, in combination with the soft excess, as the result of the reprocessing, from the inner regions of the disk, of the incident hard X-ray radiation. We adopted the model developed by \cite{coplrefl}, \textsc{coplrefl} in \textsc{Xspec}, which computes the reflection spectrum produced by a hard continuum ($\Gamma < 1$). In agreement with \cite{coplrefl}, we found that the reflection spectrum is produced by a highly ionized material, (high ionization factor: $\log(\xi) \sim 3$, z=0.037). If we estimate the radius which corresponds to the gravitational redshift obtained from the fit, we found r$ \approx 55$ km, so we ruled out that this is a measure of the gravitational redshift.
More likely this redshift could be due to the bulk motion of matter funneled by the magnetic lines at the magnetospheric radius. A simple estimate of the free-fall velocity correspondent to this radius gives a value of 0.037c, which is consistent with our best fit value. Convincing arguments in favour of reprocessing of hard X-ray radiation were given in \cite{2004ApJ...614..881H}. They interpreted the soft excess down to 0.7 keV as due to optically thick reprocessing by the accreting material.  

%
\subsubsection*{Properties of the accretion disk}
Finally, we found that an additional cold thermal component is necessary at lower energies (E$<0.6$ keV, see Fig. \ref{fig:fit_pheno}) and interpreted it as emission from a truncated accretion disk with inner disc temperature $kT_{in} = 0.065$ keV). 
The inner disk radius is located at some $10^{3}$ km, which is fairly consistent with the location of the Alfvèn radius.
The soft excess of LMC X-4 has been analyzed many times by previous authors. The cold emission, with $kT_{\mathrm{bb}} \approx 0.06$ keV was also found with the broad-band analysis of \textit{BeppoSAX} data by \cite{2001ApJ...553..375L}. The presence of a black body component was detected also by \cite{nielsen09_spec} and \cite{2019ApJS..243...29A}, with a slightly lower temperature compared to ours ($kT\approx$ 0.04 keV). However, in their modelling they do not represent the soft excess above 0.6 keV with the reflection component, but they used thermal bremsstrahlung. 
We tried to fit our data of high flux epoch with their model (fixing $N_{H}=5.708 \times 10^{20}$ cm$^{-2}$) and considering a power-law with high energy cut-off for the hard X-rays, with the addition to a gaussian component to represent the iron line (with fixed energy line and width at 6.4 keV and 0.4, respectively). The fit we obtained (with slightly different values of the model parameters) was not satisfactory (\chired=1.6 with 446 d.o.f.), at variance with the fit with a series of physical model components proposed by us. We underline that \cite{2002ApJ...579..411P} argue against thermal brehmstrahlung as source of soft emission, because of its pulsating nature. In fact, they underlined that the plasma region producing this emission would be too large to be consistent with the pulse period.\\Some authors drew different conclusions about the softer thermal emission origin. Both \cite{2010ApJ...720.1202H} and \cite{brumback2020} modelled it with a black body but they obtained a higher value than ours for the peak temperature (kT$_{\text{bb}} \sim 0.168$ keV). However, we underline that they did not incorporate a reflection component in their models (fitting the iron feature at $\sim 6.4$ keV with a gaussian) and
the discrepancy between their results with respect to ours may be due to this (compare our Fig. \ref{fig:best_bw} with Fig. 6 in \citealt{2010ApJ...720.1202H} and the second panel of Fig. 6 in \citealt{brumback2020}). 

\subsubsection*{Spectral evolution along the super-orbital period and evidence for disk precession}
After having determined the various processes which contribute to the X-ray emission, we studied how they evolve along the super-orbital modulation.
First, we found that the X-ray spectrum in the $15-60$ keV energy range do not require changes in the spectral parameters to be well fitted. The variation of the flux intensity observed in the lightcurve, at this energy range, can be reproduced throughout a scaling of the best fit model of high flux epoch with a multiplicative constant that decreases departing from $\psi_{\mathrm{sup}} \equiv \psi_{\mathrm{max}} \approx 0.5$ (see Fig. \ref{fig:onlyBAT}).\\
Figs. \ref{fig:test1} and \ref{fig:res_Kmaxmodel} show that the same is not valid for the broad-band phase-resolved spectra. A second order effect plays a role and, even if the lightcurves at different energy bands follow the same super-orbital trend (see Fig. \ref{fig:xrt_total_lcurve}), the spectra cannot be reproduced by a simple scaling of the model of high flux epoch.
Instead, we found good fits for the broad-band spectral evolution if we considered that the three main components, the cold thermal emission, the reflection and the emission by the accretion column, that are produced by different regions of the accretion flow, are left to vary independently. The resulting fits are more promising than the previous ones, as the residuals show in Fig. \ref{fig:res_freenorms}.\\
Then, we inspected how the normalizations of the \textsc{coplrefl} and \textsc{diskbb} components vary at different phases. The first and the middle panel of Fig. \ref{fig:test_final_yeah} show that they follow a similar trend, with a substantial drop at the wings of the super-orbital modulation. If our interpretation is correct, i.e. that the softest thermal emission is in fact a Shakura-Sunyaev disc, the variation of its normalization describes how the inclination of the disc changes. Under this assumption, we obtained that if
the inclination at $\psi_{\mathrm{max}}$ is $ \theta_{\psi_{\mathrm{max}}} \approx 60$\textdegree   (i.e. the observed inclination of the system), the disk inclination increases as far as $\psi_{\mathrm{sup}}$ departs from $\psi_{\mathrm{max}}$, so that it is seen edge-on at the epochs of minimum flux ($\theta_{\psi_{\mathrm{min}}} \approx 90$\textdegree ). At $\psi_{\mathrm{\mathrm{min}}}$ the radiation emitted in the vicinity of the compact object, as well as the reflection component, are covered by the disk thickness.\\
We underline that, even if we have modelled the softer part of the spectrum with a \textsc{diskbb} component, there are different interpretations in literature that must be discussed in order to give a more comprehensive view of the soft emission. 
For instance, \citealt{coplrefl}, that we cited above, modelled the softer part of the spectrum with an additional reflection component characterized by a lower ionization level ( $\log(\xi) \sim 1.5$), that is possibly located in the outer disc. We did not successfully fitted this further reflection component with our data. However, we underline that even if the thermal component modelled by us would be more phenomenological than physical, the variation of its normalization implies a real physical diminishing of the flux received, so that our final results are still valid and not strictly model dependent for what regards the softer X-ray emission.\\
With this work, performed with a dedicated monitoring campaign, we confirm the results obtained by other authors. \cite{nielsen09_spec}, with high resolution spectroscopy in one high state observation and some transition state observations, found that the variability of the emission lines produced by the accretion disk, as well as those produced within the stellar wind, can be explained in terms of precession of the accretion disk, which they concluded to be seen edge-on in the epochs of minimum flux. In addition to this, they identified a relativistic Doppler splitting of the iron line in the inner regions of the accretion disk, suggesting that these lines can originate in the inner disk warp region. This is in agreement with the results we obtained with our disk reflection study.\\
\cite{2019PASJ...71...36I} modelled the super-orbital phase-resolved 4-20 keV MAXI lightcurve of LMC X-4 with their precession model \citep{2012PASJ...64...40I} in which the outer portion of the accretion flow, that they modelled as a ring, is supposed to precess due to the tidal forces that the companion star exerts on it. As a consequence, the outer portion of the disc obscures the inner regions of the accretion flow. At first sight, their conclusions could be thought to be different from ours, as they did not find energy dependence on the X-ray modulation. However, we underline that they modelled the light curve in the 4-20 keV energy range, and are possibly referring to what we considered to be the first order effect of the disc precession, which clearly modulates both the soft and hard X-ray emission, as shown in Fig. \ref{fig:xrt_total_lcurve}. This interpretation is strengthened also if we considered that their model predicts the outer disc radius to be located at some $9\times 10^{5}$ km, which is approximately consistent with the orbital separation of LMC X-4 obtained by \cite{2015A&A...577A.130F}.\\
We also found that there is a second order effect on the super-orbital modulation that produces a mismatch between the softest and hardest X-ray emission (Fig. \ref{fig:test1}). In fact, we were able to fit the super-orbital phase resolved spectra only if all the model components are left to vary independently from the others. We interpreted this effect as due by a tilting of the disc, so that the observer sees its inner and outer regions with a different inclination.
A similar conclusion is given in \cite{brumback2020}. They studied the long-term changes of the pulse profile in the soft energy, with respect to those in the high-energy and found they are well explained in terms of reprocessing of the hard X-ray photons produced by the neutron star, by the inner regions of the precessing accretion disk (we refer to \citealt{2005ApJ...633.1064H} for an exhaustive description of the model). They also found that the precessing disc is warped so that the shape of the pulse profile can be explained with the framework proposed by \citealt{2005ApJ...633.1064H} in which the inner and outer regions of the disc are tilted one with respect to another.  

\section{Summary and Conclusions}
Super-orbital modulations are present in a high number of X-ray binaries. Understanding this type of variation is of fundamental importance to shed light on the mechanisms related to the accretion processes and to the evolution of interacting binary systems. LMC X-4 is one of those HMXBs with a long-lasting super-orbital period, which have been extensively studied. So far, a strong limit on the observational studies was given by the lack of broad-band, uniform and dense data collected along the super-orbital modulation. With the aim to improve it, we asked for a \textit{Swift} monitoring campaign along the super-orbital period. We performed a broad-band phase-resolved spectral analysis using XRT data of the campaign and BAT survey data collected during $\approx 4.75$years.\\
First, we jointly fit the 0.3-50 keV XRT and \textit{NuSTAR} spectra  collected at the epoch of high flux with a self-consistent physical model. The spectrum from 50 keV down to 2 keV is strongly dominated by the emission and scattering processes which take place within the accretion column. The emission line observed at E$\sim$6.5 keV, together with the first part of the soft excess down to 0.6 keV, are produced by the reflection of the hard X-rays by the inner regions of the accretion disk.\\ 
Finally, the softer portion of the spectrum, at E$\leq$0.6 keV is dominated by the emission of a cold thermal component that we modelled as the outer portion of the accretion disk.
We then analyzed the evolution of the various components along the super-orbital period and found
that a stronger, first order effects acts in a flux decrease both in the soft and high energy of the X-ray spectrum. Moreover, a second order effect, that we addressed to a tilting of the disc, plays a role and produces a mismatch between the soft and hard X-ray emission .

\section*{Acknowledgements}
The Authors thank the anonymous Referee for Their comments and suggestions that improve the quality of the paper.
EA, ADI and GC acknowledges funding from the Italian Space Agency, contract ASI/INAF n. I/004/11/4.
EA, ADI, MDS, AS and GC acknowledge financial contribution from the agreement ASI-INAF n.2017-14-H.0 and the INAF main-stream grant. 
EA thanks Milvia Capalbi for useful discussion on the XRT data analysis. RA thanks the CNES for their support.

\section*{Data availability}
The data underlying this article are available at the HEASARC public archive, as explained in Sec. \ref{sec:data}



\bibliographystyle{mnras}
\bibliography{lmc_x4.bib}




\appendix
\section{Some useful formulas}
\label{app:Characateristics of the accretion column}
In the following, we present the equations (BW07), through which we inferred the properties of the accretion column.
Results are shown in Tab. \ref{tab:bw_phys}.\\
Mass accretion flux (g s$^{-1}$ cm$^{-2}$):
\begin{equation}
    J  = \frac{\dot{M}}{\pi r_{0}^{2}},
\end{equation}
where \Mdot is the mass accretion rate and $r_{0}$ is the polar cap radius.
Mound temperature (K):
\begin{equation}
   \mathrm{ T_{th} } = 2.32\times 10^{3} \dot{M}^{2/5}r_{0}^{-2/3}
\end{equation}
Inflow speed at the mound surface (cm s$^{-1}$)
\begin{equation}
\mathrm{v_{th}}= 7.86 \times 10^{10} \dot{M} r_{0}^{-3/2} T_{th}^{-7/4}
\end{equation}
Top of the thermal mound (cm):
\begin{equation}
   \mathrm{ z_{th} } = 5.44 \times 10^{15} \frac{\dot{M} R_{*}}{M_{*} r_{0} T_{th}^{7/2} \xi \sigma_{\parallel}}
\end{equation}
Here, $\xi$ is a dimensionless parameter which quantifies the efficiency of the the escape photons in removing the kinetic energy of the gas, that has to set at rest on the surface of the thermal mound. It is defined by the following expression:
\begin{equation}
    \xi \equiv \frac{\pi r_{0} m_{p} c}{\dot{M} (\sigma_{\parallel} \sigma_{\bot})^{1/2}}
\end{equation}
$\sigma_{\parallel}$ represents the mean cross section of scattering for photons which propagate parallel to the magnetic fiels, and $\sigma_{\bot}$ the one for photons which propagate on the direction perpendicular to the magnetic field.\\
Altitude for photon trapping (cm): 
\begin{equation}
   \mathrm{ z_{trap} }= \frac{\pi r_{0}^{2} c m_{p}}{2\dot{M} \sigma _{\parallel}} = \frac{r_{0} \xi }{2} \bigg ( \frac{\sigma _{\bot}}{\sigma _{\parallel}} \bigg) ^{1/2}
\end{equation}
Altitude at the sonic point (cm): 
\begin{equation}\label{eq:zsp}
    \mathrm{z_{sp}} = \frac{r_{0}}{2 \sqrt{3}} \bigg( \frac{\sigma_{\bot}}{\sigma_{\parallel}} \bigg)^{1/2} \ln{ \frac{7}{3}}
\end{equation}
Critical luminosity (\unilum):
\begin{equation}
 \mathrm{ L_{crit} } = \frac{2.72\times 10^{37} \sigma{\bot} }{\sqrt{\sigma _{\bot} \sigma_{\parallel}}} \bigg( \frac{M_{*}}{M_{\odot}} \bigg) \bigg( \frac{r_{0}}{R_{*}} \bigg)
\end{equation}
\section{Consistency of the joint fit of the \textit{Swift}/XRT and \textit{NuSTAR} data}
\label{app:useful_figures}
\begin{figure}
\centering
\includegraphics[width=\columnwidth]{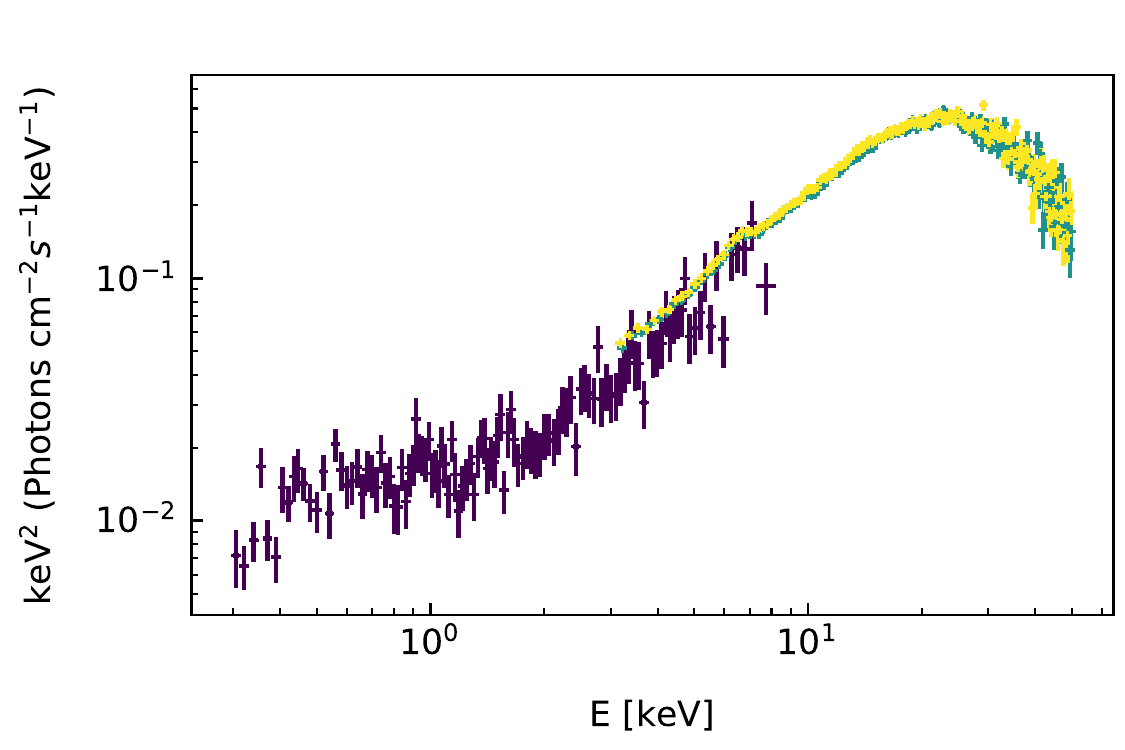}
\caption{XRT+\textit{NuSTAR} joint spectra unfolded through a power-law model with photon index = 0. This figures states the feasibility of jointly fit these two non simultanous spectra (see the introduction of Sec. \ref{sec:broad-band}).}\label{fig:folded_jointfit}
\end{figure}

In our analysis we used non-simultaneous \textit{Swift}/XRT and \textit{NuSTAR} data to determine the spectrum at the epoch of maximum flux. The feasibility of the joint fit of these data is shown in
Fig. \ref{fig:folded_jointfit}.

\bsp	
\label{lastpage}
\end{document}